# Bulk nanocrystalline Al alloys with hierarchical reinforcement structures via grain boundary segregation and complexion formation

Tianjiao Lei [1], Jungho Shin [2], Daniel S. Gianola [2], Timothy J. Rupert [1,*]

[1] Department of Materials Science and Engineering, University of California, Irvine, CA 92697, USA

[2] Materials Department, University of California, Santa Barbara, CA 93106, USA

* Email: trupert@uci.edu

## Abstract

Grain size engineering, particularly reducing grain size into the nanocrystalline regime, offers a promising pathway to further improve the strength-to-weight ratio of Al alloys. Unfortunately, the fabrication of nanocrystalline metals often requires non-equilibrium processing routes, which typically limit the specimen size and require large energy budgets. In this study, multiple dopant elements in ternary Al alloys are deliberately selected to enable segregation to the grain boundary region and promote the formation of amorphous complexions. Three different fully dense bulk nanocrystalline Al alloys (Al-Mg-Y, Al-Fe-Y, and Al-Ni-Y) with small grain sizes were successfully fabricated using a simple powder metallurgy approach, with full densification connected directly to the onset of amorphous complexion formation. All the compositions demonstrate densities above 99% with grain sizes of <60 nm following consolidation via hot pressing at 585 °C. The very fine grain structure results in excellent mechanical properties, as evidenced by nanoindentation hardness values in the range of 2.2-2.8 GPa. Detailed microstructural characterization verifies the segregation of all dopant species to grain boundaries as well as the formation of amorphous complexions, which suggests their influential role in aiding effective consolidation and endowing thermal stability in the alloys. Moreover, nanorods with a core-shell structure are also observed at the grain boundaries, which likely contribute to the stabilization of the grain structure while also strengthening the materials. Finally, intermetallic particles with sizes of hundreds of nanometers form in all systems. As a whole, the results presented here demonstrate a general alloy design strategy of segregation and boundary evolution pathway that enables the fabrication of multiple nanocrystalline Al alloys with hierarchical microstructures and improved performance.

## Keywords

# 1. Introduction

Al alloys are a class of structural materials widely used in aerospace and gaining growing interest for automotive applications because of a combination of exceptional strength-to-weight ratio, high stiffness, and superior specific strength [1,2,3]. Conventional Al alloys typically have grain sizes in the micrometer range, and common alloying elements include Zn, Mg, and Cu. These alloying elements generally promote the formation of secondary phases to strengthen the materials [4,5], leading to yield strengths that can sometimes exceed 700 MPa. To further improve the strength, grain refinement has been identified as a promising route since the strength-to-weight ratio can be increased without requiring more alloying elements, which are often heavier than the base Al. For example, Ma et al. [6] showed that, compared to an as-extruded coarse-grained 7075 alloy with a yield strength of 283 MPa, the yield strength of the same alloy composition but with an ultra-fine grain size is 583 MPa. With grain refinement, grain boundary strengthening is the predominant mechanism as predicted by the Hall-Petch relationship. Consequently, further refinement of grain sizes down to the nanoscale regime may offer a chance for even better mechanical properties. Li et al. [7] sputter deposited columnar nanotwinned Al-Fe alloy films with an average grain size of ~5 nm and obtained a yield strength exceeding 1.8 GPa, as measured through compression tests on micron-sized pillars. However, despite the promise of outstanding properties, the high density of grain boundaries in nanocrystalline metals and alloys also typically results in poor thermal stability so that undesired grain growth can easily occur during material processing or in service. Given the low melting point of Al and its alloys, even modest temperature exposures can lead to deleterious microstructural evolution. For example, Ahn et al. [8] investigated the effect of degassing temperature on the microstructure of a nanocrystalline Al 5083 alloy and observed an increase in grain size from 50 nm to 118 nm after degassing at 500 °C for 2

h. Moreover, the grains further coarsened to a mean size of 214 nm after cold isostatic pressing and forging processes. As a result, synthesis of nanocrystalline alloys usually requires far-from-equilibrium processing routes, such as magnetron sputtering, electrodeposition, and pulsed laser deposition techniques, which commonly lead to specimens with small overall dimensions on the micron scale or smaller [9]. For instance, Devaraj et al. [10] employed magnetron sputtering to fabricate coarsening-resistant Al-Mg thin films, yet these materials only had a thickness of ~100 nm. In contrast, powder metallurgy methods can be easily scaled up and should require much less energy compared to techniques such as high-pressure torsion or equal channel angular pressing, which necessitate multiple plastic deformation cycles, each requiring high applied forces.

Evidence is building in the literature to support the idea that the most effective method for stabilizing a nanocrystalline grain structure is the addition of dopant elements that can segregate to grain boundaries, which reduces the driving force for grain growth and/or provides kinetic stabilization on the grain structure. The grain boundary energy for a dilute solution is [11]:

$$\gamma = \gamma_0 - \Gamma(\Delta H^{seg} + kT\log X), \quad (1)$$

where $\gamma_0$ is the grain boundary energy of the pure material, $\Gamma$ is the dopant excess at grain boundaries, $\Delta H^{seg}$ is the segregation enthalpy, $kT$ is the thermal energy, and $X$ is the global dopant content. In order for dopant atoms segregating to and stabilizing grain boundaries, appropriate dopant elements need to be chosen, with previous studies providing guidelines such as large atomic size mismatch with the solvent [12] and, correspondingly, low bulk solubility [13]. Chookajorn et al. [14] built on this idea to create a theoretical framework based on a thermodynamic model to evaluate and design stable binary nanostructured alloys. This model considers two key thermodynamic parameters, which are the enthalpy of mixing in the crystalline state for the grain interior and the dilute-limit enthalpy of segregation for the grain boundaries.

These authors then identified a stable nanostructured alloy system, W-Ti, which was shown to retain a 20 nm grain size even after annealing at 1100 °C for one week, owing to the segregation of Ti to grain boundaries and consequent stabilization. Later work by Murdoch and Schuh [15] employed a Miedema-type model to estimate the grain boundary segregation enthalpy in binary alloy systems, where positive and negative enthalpy values indicated segregation and depletion of dopant atoms at grain boundaries, respectively. The enthalpy calculation was expanded to ~2500 binary alloys, which can serve as a rapid screen tool for the stability of nanostructured alloys. It is worth noting that grain boundary segregants can concurrently provide a kinetic drag term that slows grain growth [16]. As a whole, alloying with segregating dopants is generally being accepted as the key process necessary for the creation of thermally-stable nanocrystalline alloys. However, the vast majority of previous work has focused on binary systems where only a single segregating species was added. Recent evidence suggests complex segregating behavior in multi-component systems is ultimately driven by the nature of the thermodynamic interactions between multiple segregating species, the kinetic ability to diffuse to the boundaries, and the nature of site competition at various boundaries [17,18,19,20].

Segregation of dopant atoms can not only alter the chemistry of grain boundaries but also change the local equilibrium structure of these interfaces. A grain boundary can be described as a phase-like entity, known as a *complexion*, and may transform between different structures as chemistry and/or temperature changes [21, 22]. One type of complexion is an amorphous intergranular film, which occurs when a grain boundary undergoes a premelting transition, locally melting below the bulk melting point of the alloy and yet retaining an equilibrium thickness at the boundary. Amorphous complexions have been shown to significantly enhance thermal stability of nanocrystalline alloys, as Grigorian and Rupert [23] demonstrated that Cu-Zr-Hf alloys with thick amorphous complexions exhibit nanosized grains even after two weeks of annealing at a temperature higher than 95% of the solidus temperature. In addition to improving thermal stability, amorphous complexions have been associated with enhanced diffusion and the phenomenon of activated sintering [24]. For example, in a W specimen that was doped with Ni, Gupta et al. [25] observed the presence of nanoscale amorphous Ni-enriched films at grain boundaries well below the bulk eutectic temperature, which gave rise to enhanced mass transport and consequently solid-state activated sintering. In a different system, Donaldson and Rupert [26] reported that the formation of amorphous complexions in a Cu-4 at.% Zr alloy consolidated from ball-milled powders coincided with a significant increase in the density of bulk samples.

In the present study, our goal is to choose alloying elements that can segregate to grain boundaries and form amorphous complexions so that bulk Al-rich alloys with both nanosized grains and high density can be achieved. Three elemental combinations are selected: Al-Mg-Y, Al-Fe-Y, and Al-Ni-Y. Ternary systems are targeted due to recent studies which provide evidence of better thermal stability and thicker amorphous complexions, along with their much slower critical cooling rate needed to retain amorphous complexions compared to binary systems [23,27]. All of the compositions were successfully fabricated into fully dense bulk pellets with a diameter of ~1.4 cm and height of ~1 cm using a simple and scalable powder metallurgy approach. Moreover, the bulk samples showed full densification to >99% density and retained a grain size of ~50 nm under proper hot pressing conditions. Microstructural characterization reveals a hierarchical structure containing a nanocrystalline Al-rich grain structure and additional reinforcing phases with different structures and characteristic length scales. On a very fine scale, complete segregation of dopant atoms to grain boundary regions and formation of amorphous grain boundary complexions were observed, with the latter feature being connected to activated sintering and mechanical behavior. Nanorods with a core-shell structure were observed to nucleate at grain boundaries as well, providing a reinforcing phase on the nanometer scale. The internal structure of the nanorods is amorphous and clear partitioning of Y to an outer shell layer is observed. On a larger length scale, intermetallic particles with sizes of a few hundred nanometers were uniformly distributed within the microstructure. The end result is that the Al alloys reported here readily consolidate from powder form into bulk samples and possess a hierarchical nanostructure that both stabilizes the grains and results in high mechanical strength, all of which appear to be consequences of the formation of segregation-mediated complexion formation. The present study offers a simple approach to synthesize fully dense bulk nanocrystalline Al alloys via engineered grain boundaries.

## 2. Materials and methods
## 2.1 Materials selection

In the present study, the goal is to fabricate bulk nanocrystalline Al-rich ternary alloys with both full density and small grain size, by employing grain boundary segregation of dopant elements expected to stabilize the grain size and promote the formation of amorphous complexions. Therefore, appropriate selection of dopant species is a crucial step. First, positive segregation enthalpies are desired indicating that dopant atoms tend to segregate to grain boundaries. In addition, limited solubility is beneficial as another sign that most of the dopant atoms will not dissolve in the Al matrix and instead will segregate to grain boundaries. Next, pair-wise enthalpy of mixing should be negative so that atoms of different elements are likely to mix with each other, as this has been shown to be critical for the formation of amorphous complexions as proposed by Schuler and Rupert [28]. Finally, large atomic size mismatch is another important factor in analogy to bulk metallic glasses, as a significant difference in atomic sizes among the constituent elements has been shown to be beneficial for a larger extent of dense randomly packed atomic configurations [29]. Due to the similar structural disorder in both metallic glasses and amorphous complexions, the criterion for metallic glasses can help serve as a guide for amorphous complexion design. Based on the above factors, three alloy systems of Al-Mg-Y, Al-Fe-Y, and Al-Ni-Y were chosen, with roughly 2 at.% used for each dopant element. Each system includes one alkaline-earth/transition metal and one rare earth metal to maximize the atomic size mismatch [30,31], while the selection of Mg, Fe, and Ni allows for a range of common metals to be studied. In addition, all dopant elements have limited solubility in Al [32,33,34,35], and Mg, Fe, and Ni have been shown to segregate to grain boundaries in Al alloys [36,37,38]. Although the segregation of Y to grain boundaries in Al has not been reported directly to date, grain boundary segregation of a similar rare-earth element, Ce, and complexion formation were observed in a ternary nanocrystalline Al alloy [36]. Computational studies have shown that pair-wise mixing enthalpies are negative in all of the systems [39,40,41].

**2.2 Materials fabrication**

To synthesize the nanocrystalline alloys, powders of elemental Al (Alfa Aesar, 99.97%, - 100+325 mesh), Mg (Alfa Aesar, 99.8%, -325 mesh), Fe (Alfa Aesar, 99.9%, -20 mesh), Ni (Alfa Aesar, 99.9%, APS 2.2-3.0 micron), and Y (Alfa Aesar, 99.6%, -40 mesh) were first ball milled for 10 h in a SPEX SamplePrep 8000M high-energy ball mill using a hardened steel vial and milling media. All milling processes were conducted in a glovebox filled with Ar gas and with an oxygen level lower than 0.05 ppm, in order to avoid oxidation during milling. A ball-to-powder weight ratio of 10:1 was used with 3 wt.% stearic acid as a process control agent to prevent excessive cold welding. As an initial test of each alloy's thermal stability, the as-milled powders were encapsulated under vacuum in quartz ampules and annealed at either 270 °C or 540 °C for 1 h in a tube furnace (MTI International GSL-1100X-NT). Each sample was then dropped into a water reservoir to quench as fast as possible.

For fabrication of bulk specimens, the as-milled powders were loaded into a ~14 mm inner diameter graphite die and then consolidated into cylindrical bulk pellets using an MTI Corporation OTF-1200X-VHP3 hot press. This system is comprised of a vertical tube furnace with a vacuum-sealed quartz tube and a hydraulic press. The powders were first cold pressed for 10 min under 100 MPa at room temperature to form a green body, and then hot pressed for 1 h under 100 MPa at different temperatures to enable densification, grain boundary segregation, and complexion formation. The hot pressing temperatures ($T_{HP}$) used were 449 °C, 495 °C, 540 °C, and 585 °C, corresponding to $0.77T_m$, $0.82T_m$, $0.87T_m$, and $0.92T_m$, respectively, where $T_m$ = 663 °C is the melting temperature of pure Al [42]. The heating rate to reach the target $T_{HP}$ was 10 °C/min. Finally, the pellets were naturally cooled down to room temperature, which typically took more than 4 h. For the investigation of microstructural features representative of the high temperature state, one Al-Ni-Y pellet that was originally hot pressed at 540 °C and then naturally cooled down to room temperature was subsequently annealed again at 540 °C for 10 min. This specimen was then very quickly quenched by being placed on an Al heat sink sitting in liquid nitrogen, so that the high temperature structure can be frozen in and studied. For the remainder of the paper, the hot-pressed and then naturally cooled pellets will be referred to as *naturally cooled samples*, while the annealed and subsequently quenched sample will be referred to as the *quenched sample*.

## 2.3 Materials characterization

The cylindrical pellets were cut into two equal semi cylinders using a low-speed diamond saw so that the cross-sectional plane could be studied. All cross-sectional surfaces were first ground with SiC grinding paper down to 1200 grit and then polished with monocrystalline diamond pastes down to 0.25 μm. For both powders and bulk pellets, X-ray diffraction (XRD) scans were collected using a Rigaku Ultima III X-ray diffractometer with a Cu Kα radiation source operated at 40 kV and 30 mA and using a one-dimensional D/teX Ultra detector. Phase identification, grain size analysis, and weight fraction calculation from XRD scans were all carried out using Rigaku PDXL software, which is an integrated powder X-ray analysis software package. Grain size measurements were performed using the Halder-Wagner method [43] with a $LaB_6$ calibration file used as an external standard, and weight fraction calculation were performed via Rietveld refinement [44].

The density of each pellet was determined by imaging polished surfaces with a scanning electron microscope (SEM), following ASTM standard, E 1245-03 [45]. For each sample, at least 30 images of different areas were taken for the density calculation and ImageJ software [46] was used for image processing to obtain the density for each image. The final density of each sample is an average of the values obtained from different SEM images, and error bars are standard deviations from these measurements. SEM imaging, energy-dispersive spectroscopy (EDS) measurements, and back-scattering electron imaging were performed in an FEI Quanta 3D FEG dual-beam SEM/Focused Ion Beam (FIB) microscope. Conventional and scanning transmission electron microscopy ((S)TEM) paired with EDS were used to examine the grain size and elemental distribution inside of a JEOL JEM-2800 S/TEM, which was operated at 200 kV and equipped with a Gatan OneView IS camera and two dual dry solid-state 100 $mm^2$ EDS detectors. TEM samples were fabricated by using the FIB lift-out method [47] with a Ga$^+$ ion beam in the FEI Quanta 3D FEG dual-beam SEM/FIB microscope equipped with an OmniProbe. A final polish at 5 kV and 48 pA was used to minimize the ion beam damage to the TEM sample.

Mechanical properties of the bulk alloys were investigated by nanoindentation experiments using a Nano Indenter G200 (Agilent Technologies). For each sample, at least 50 indentations were performed with a maximum penetration depth of 400 nm and constant indentation strain rate of 0.05 s$^{-1}$. The distance between the two nearest indents was 12 μm, which was approximately 30 times the penetration depth so that interference can be avoided [48]. The final hardness value presented is an average of all tests for each sample, and the error bars represent the standard deviation.

## 3. Results and discussion

### 3.1 Sample density and thermal stability

The chemistry and structure of the as-milled powders were first characterized, as well as the general stability of these alloy powders against grain growth. Table 1 lists the dopant concentration measured by SEM-EDS for each system and the grain size of the Al-rich face-centered cubic (FCC) phase obtained from XRD for both as-milled and annealed powders. For the as-milled powders, Al-Fe-Y and Al-Ni-Y exhibit similar grain sizes of 11 and 10 nm, respectively, which are roughly half that measured for Al-Mg-Y (19 nm). After 1 h annealing at an intermediate temperature of 270 °C (0.58$T_m$), the measured grain size of all alloy systems remains unchanged, indicating no grain growth at this temperature. When the as-milled powders were annealed at a higher temperature of 540 °C (0.87$T_m$), modest grain growth was observed. Namely, the Al-rich FCC grains in Al-Fe-Y and Al-Ni-Y both grow to 34 nm, while those in Al-Mg-Y increase to 39 nm. Therefore, all three alloy systems show very small grain sizes for the as-milled powders and thermal stability that is comparable or higher than other studies. For example, Hanna et al. [49] reported that cryomilled Al-5083 alloy powders that were stabilized using diamantane molecular diamonds exhibited an average grain size of 22 nm in the as-milled condition, which is twice as large as the as-milled Al-Fe-Y and Al-Ni-Y powders here. The grain size of the Al-5083 powders roughly remained the same or increased to ~52 nm after 1 h annealing at 300 ºC or 500 ºC, respectively, similar to the trend observed here but indicating more growth.

The nanocrystalline powders were next consolidated into pellets at different $T_{HP}$. Four $T_{HP}$ values (449 ºC, 495 ºC, 540 ºC, and 585 ºC) were investigated, and the corresponding pellets were characterized by various techniques as detailed below. Figure 1(a)-(c) presents low-magnification cross-sectional SEM images with the inset value being the corresponding pellet density. For the lowest $T_{HP}$ (449 ºC), the pellet densities of Al-Fe-Y and Al-Ni-Y are very low (less than 40%) as there are few areas consolidated well (darker contrast) on the sample surface. In contrast, the pellet density of Al-Mg-Y is much higher at ~90%. When $T_{HP}$ is raised to 495 ºC, the density of Al-Mg-Y increases to 94.1%, while those of Al-Fe-Y and Al-Ni-Y show a significant improvement from their previously low values, reaching 95.1% and 93.8%, respectively. The substantial increase in the density of consolidated pellets from $T_{HP}$ = 449 to 495 ºC is an example of activated sintering that we hypothesize can be attributed to amorphous complexions forming and allowing for enhanced diffusion. For example, Luo and Shi [50] performed computational studies and predicted that the amorphous complexions can form in a temperature range of 0.6-0.85 of the bulk solidus temperature. Subsequently, Luo and colleagues [25,51,52] confirmed these predictions by using high resolution TEM to reveal the existence of amorphous complexions well below the bulk eutectic temperature in various systems, which are responsible for activated sintering. In another study, Donaldson and Rupert [26] reported a dramatic increase in the density at 850 °C (~0.83 of the melting temperature of pure Cu) for a Cu-4 at.% Zr alloy, which coincides with the formation of amorphous complexions as observed using high resolution TEM. In the present study, the significant increase in pellet density for both Al-Fe-Y and Al-Ni-Y also occurs within a homologous temperature range from $0.77T_m$ to $0.82T_m$, and is most likely associated with amorphous complexion formation, a hypothesis which will be tested directly in Section 3.2. The lower onset temperature for activated sintering in the Al-Mg-Y system suggests the important role of alkaline-earth/transition metal element choice in controlling density evolution during consolidation. With further increase in pressing temperature, all three systems show similar values and reach fully dense (with a density > 99%) by $T_{HP}$ = 585 °C.

For better visualization of the pellet density evolution, density as a function of $T_{HP}$ is presented in Figure 1(d). From $T_{HP}$ = 449 °C to 585 °C, the density of Al-Mg-Y increases approximately linearly with a slope of ~ 0.07%/°C, while the density of Al-Fe-Y and Al-Ni-Y more than doubles from 449 °C to 495 °C and then follows a similar trend as that of Al-Mg-Y at higher $T_{HP}$. To check whether the Al-Mg-Y system shows a significant density increase for a certain $T_{HP}$ range, a lower $T_{HP}$ of 400 °C was investigated and its corresponding pellet density is also shown Figure 1(d). The pellet density of Al-Mg-Y with $T_{HP}$ = 400 °C is 38.3%, similar to those of Al-Fe-Y and Al-Ni-Y using $T_{HP}$ = 449 °C. Therefore, the Al-Mg-Y system also exhibits a dramatic density increase within a certain $T_{HP}$ range but at a lower hot pressing temperature than Al-Fe-Y and Al-Ni-Y. This indicates the importance of the alkaline-earth/transition metal element in controlling the chemical and structural nature of the grain boundary region, and ultimately in governing consolidation behavior.

After verifying that all systems can be consolidated into fully dense bulk pieces, the next step is to evaluate the resulting grain size and its evolution during consolidation. Figure 2 presents the XRD grain size of the Al-rich FCC phase as a function of $T_{HP}$, which shows that for all three alloys, the size of the matrix grains increases monotonically with increasing $T_{HP}$. The smallest average grain sizes are in the range of ~20-30 nm, while the largest average grain size values are ~45-55 nm. Among the three systems, Al-Mg-Y exhibits a slightly larger grain size than the other two alloys for all temperatures. However, the fact that all three systems still retain relatively small nanocrystalline grain sizes even after hot pressing at $0.92T_m$ signals exceptional thermal stability of the alloy systems in their bulk form. Due to the large driving force for grain coarsening, nanocrystalline alloys that exhibit nanosized grains in a powder form tend to grow into the ultrafine regime along the processing pathway to bulk form that involves high temperature and/or pressure. For instance, Sasaki et al. [53] employed mechanical alloying and spark plasma sintering under a load of 35 kN at 480 °C to fabricate a bulk Al-Fe alloy. In this study, whereas the average size of the Al-rich FCC grains of the as-milled powders was 26 nm, the sintering process caused some of the Al grains to coarsen to the 0.5-1 μm regime, although some nanoscale grains still persisted with a mean size of 76 nm. In the present study, all of the ternary alloys investigated maintain their nanostructure throughout all stages of processing, from powder milling to consolidation of bulk samples, which is presumably a direct consequence of the segregation-mediated grain boundary states.

TEM was next used to confirm the nanocrystalline grain structures in a subset of samples. Figure 3(a) presents representative bright field TEM micrographs for the highest $T_{HP}$ along with the cross-sectional optical images of the corresponding bulk pieces. The TEM images confirm the nanocrystalline nature of the alloys and show equiaxed grain shapes. The cross-sectional optical images present the bulk size of the pellets, with diameters of ~1.4 cm and heights of ~0.5-0.7 cm after polishing. Figure 3(b) shows two optical images of one as-consolidated pellet before polishing, the height of which is ~1.3 cm. TEM measurements of grain sizes were performed by using both bright-field TEM and high-angle annular dark-field (HAADF) STEM micrographs, with care taken to exclude intermetallic grains. For each pellet, at least 100 FCC grains were measured, and the cumulative distribution functions for grain sizes in each alloy are shown in Figure 3(c). All systems exhibit grain sizes that range from approximately 20 to 100 nm. The average grain sizes of the Al-rich FCC phase for Al-Mg-Y, Al-Fe-Y, and Al-Ni-Y, are 58±19 nm, 54±17 nm, and 57±16 nm, respectively, all of which are close to the values obtained from XRD. The grain size distributions are all well-described by a log-normal distribution, meaning that no abnormal grain growth was detected.

Overall, Figures 2 and 3 show that all three alloys were successfully fabricated into bulk pieces with both small grain size (<60 nm) and high density (>99%). Figure 4 shows a comparison of the grain size and pellet density between the Al alloys created in this study and other Al-rich alloys from the literature, with specific examples discussed below. Le et al. [54] used spark plasma sintering to fabricate fully recrystallized Al samples with a diameter of 20 mm and a height of approximately 4 mm. The size of the smallest grain was 0.8 μm and the sample density is ~99%. Furukawa et al. [55] employed equal-channel angular pressing and torsion straining separately to a bulk Al-Mg solid solution alloy, which resulted in average grain sizes of 200 nm and 90 nm, respectively. By consolidating chemically synthesized nanocrystalline Al powders, Haber and Buhro [56] obtained samples with various density and grain size combination, among which the highest density of 95% had a corresponding grain size of 200 nm. Topping et al. [57] performed quasi-isostatic forging on cryomilled Al-Mg powders and obtained fully dense disks that were ~20 mm thick and 250 mm in diameter. The average grain sizes were 194 nm, 259 nm, and 336 nm for forging temperatures of 250 ºC, 350 ºC, and 450 ºC, respectively. Compared to the studies discussed above where achieving grain sizes below 100 nm was elusive, the present work shows bulk pieces with a combination of much smaller grain size (<60 nm) and higher density (>99%). Moreover, a much simpler processing route that requires less equipment (only a high-energy ball mill and a hot press) and energy, and is innately scalable, is employed. For example, it requires much less energy and pressure to hot press versus torsion straining under a pressure of ~5 GPa to a torsional strain of ~7, as reported in Ref. [55]. The key to the exceptional combination of density and grain size is the alloy selection and judicious processing pathways that promote grain boundary enrichment and the formation of amorphous complexions, which suggest that simple yet effective processing route can be adopted by choosing proper dopants (Section 2.1). To verify the existence of these interfacial structures, the local grain boundary environment is investigated in the next section.

**3.2 Grain boundary chemistry and structure**

Although all dopant species were chosen because they were expected to segregate to grain boundaries, this hypothesis is tested by investigating the spatial distribution of dopants. The pellets with the lowest $T_{HP}$ were first selected for study, before moving on to higher $T_{HP}$. Figure 5 shows HAADF STEM images corresponding to the lowest $T_{HP}$, along with elemental maps from the same region. The grain boundaries can be identified in the HAADF STEM image itself and corroborate previously measured grain sizes, but additional confirmation of their locations is provided by the elemental map of Ga (introduced during the preparation of TEM samples), as Ga atoms are known to segregate to grain boundaries in Al alloys [58]. For each system, the distribution of dopant atoms is not uniform, and in all three system both dopant species tend to segregate to the same regions where the concentration of Ga is elevated. Therefore, we can confirm that co-segregation occurs in each alloy system. Experimentally, Pun et al. [37] observed that Mg atoms depleted in the crystal interior and enriched at grain boundaries in ball milled nanocrystalline Al-Mg powders, while Balbus et al. [17] reported segregation of both Ni and Ce atoms to grain boundaries in a sputtered nanocrystalline Al-Ni-Ce film. Computationally, Babicheva et al. [59] performed molecular dynamics simulations and showed grain boundary segregation of Fe in a nanocrystalline Al-rich binary alloy. In the present work, even though all dopant species segregate to grain boundaries, the extent to which each element segregates varies. Y strongly segregates to the grain boundaries in all systems and Mg appears to segregate more strongly than Fe and Ni. The strong segregation of Y is reasonable since the atomic radius of Y atoms is much larger than those of all other elements [60], meaning that Y atoms will prefer locations with a higher free volume such as the grain boundary region. The stronger segregation tendency of Mg than the other two transition metal elements is consistent with the calculations of Murdoch and Schuh [15] showing that the segregation enthalpy of Mg in Al is more positive than that of Fe or Ni in Al, and therefore Mg should be a stronger segregant in an Al matrix phase. In addition, the pair-wise mixing enthalpy between Mg and Y is negative (Table 2), suggesting that Mg atoms also like to be next to Y atoms. These synergistic effects in segregation and co-segregation tendencies for the alkaline-earth and transition metal elements in the ternary Al alloys could provide an explanation for the earlier onset temperature for activated sintering in Al-Mg-Y as compared with Al-Fe-Y and Al-Ni-Y.

In the present study, the dopant species were selected to promote grain boundary segregation in service of forming amorphous complexions once the boundaries were chemically enriched. Therefore, the next step is to examine the grain boundary structure in detail, and the Al-Ni-Y pellets are chosen for more in-depth inspection. From Figure 1, the temperature required for amorphous complexion formation for Al-Ni-Y is hypothesized to be between 449 °C to 495 °C, corresponding to a homologous temperature range between $0.77T_m$ to $0.82T_m$, as the density increased more than two orders of magnitude within this regime, so two $T_{HP}$ values above the observation of activated sintering were selected (495 and 540 °C). Amorphous complexions were indeed observed in naturally cooled samples consolidated at these temperatures, with these features outlined by dashed lines in Figures 6(a) and (b) showing two representative amorphous complexions for each $T_{HP}$. Care was taken to ensure that the grain boundary region was viewed in an edge-on condition by confirming that the complexion thickness does not vary in both under- and over-focused imaging conditions [61]. These images clearly demonstrate the disordered nature of the complexions and the crystalline structure of grains (denoted "G1" and "G2" in Figure 6) adjacent to the complexions. The thickness of the amorphous complexions shown here is ~2-3 nm, although variations from boundary-to-boundary have been reported in previous work [23,27,62] and would be expected in our alloys as well. However, a detailed description of the thickness distribution is not the focus of the present study.

The existence of the amorphous complexions in the naturally cooled sample (cooled over ~4 h) further verifies the robust alloy selections in the present study, as fast quenching is typically required to kinetically freeze the high-temperature interfacial structures [62]. Although amorphous complexions are stable at high temperatures, they would be metastable at room temperature and therefore could transform back to an ordered state during a slow cooling process, similar to the crystallization behavior of metallic glasses. In the metallic glass community, early fabrication techniques involve rapid solidification methods with characteristic cooling rates in the range of $10^3$-$10^6$ K/s so that crystallization could be avoided. The resistance to nucleation and growth of crystalline phases in an undercooled melt has been related to the glass forming ability of the materials by Turnbull [63], and metallic glasses with good glass forming ability can remain amorphous even under very low cooling rates. Examples of good metallic glasses include $Zr_{41.2}Ti_{13.8}Cu_{12.5}Ni_{10.0}Be_{22.5}$ with a critical cooling rate of 10 K/s or less [64], $Mg_{65}Cu_{25}Gd_{10}$ with a critical cooling rate of ~1 K/s [65], and $Pd_{43}Ni_{10}Cu_{43}P_{20}$ with a critical cooling rate of 0.4 K/s [66]. The finding of relatively stable amorphous complexions in the present Al-rich ternary alloy systems is consistent with the prior report of Grigorian and Rupert [27], where the critical cooling rate for amorphous-to-ordered complexion transition in a ternary (Cu-Zr-Hf) Cu-rich alloys was found to be much lower than that for a comparable binary (Cu-Zr) alloy. In the present study, the critical cooling rate for the Al-rich ternary systems is evidently even slower than that for the Cu-Zr-Hf in Ref. [27] since amorphous complexions still exist within the Al alloys even after a cooling rate of 0.1 K/s. Similar observation has also been made in an Al-Ni-Ce nanocrystalline alloy, where amorphous complexions still remained after slowly cooling over the course of 2-3 h from annealing at 200 °C, 325 °C, or 380 °C [17]. Compared to the Cu-Zr-Hf alloy, where the dopant atoms had similar sizes, the increased stability is likely due to the large atomic mismatch for all elements in the present Al alloys as well as the materials probed in Ref. [17]. Therefore, proper selections of microalloying elements may not only improve the material properties while retaining the amorphous complexions, but also reduce the fabrication cost since expensive quenching steps can be avoided to keep the amorphous complexions.

Since amorphous complexions are high-temperature structures and their thickness generally tends to decrease during slow cooling, the complexions shown in Figure 6 could have been thicker under the high-temperature consolidation and then shrank during the naturally cooling process. To test this hypothesis, one Al-Ni-Y pellet that was made with $T_{HP}$ = 540 °C and naturally cooled down to room temperature was annealed again at 540 °C for 10 min. Subsequently, the pellet was placed on an Al cooling block as soon as possible with the bottom part of the block submerged in a large reservoir of liquid nitrogen, as shown in Figure 7(a). Therefore, the surface contacting the Al block was very rapidly quenched, freezing in any structures that were in equilibrium at high temperature. Figures 7(b) and (c) show high resolution TEM images of two representative amorphous complexions taken from this rapidly quenched sample. Compared to the naturally cooled sample with the same $T_{HP}$ shown in Figure 6(b), the quenched sample exhibits much thicker amorphous complexions. The two examples shown in Figures 7(b) and (c) are ~9 and ~7 nm thick, respectively, and the 9 nm is on the order of the thickest example as observed in a Cu-Zr-Hf alloy [27]. The observation of thicker complexions after quenching confirms that they evolve during slow cooling [67].

### 3.3 Multi-scale precipitates

The amorphous complexions observed in the alloys are restricted to nanoscale-thickness films along the grain boundaries. Inspection with TEM shows that additional microstructural features with nanoscale dimensions and elongated shapes also exist, which will be termed nanorods here. Above a critical $T_{HP}$, nanorods begin to form in all alloy systems. The evolution of these nanorods in the Al-Ni-Y system is shown in Figure 8 as a representative example of the three alloys, as similar observations were found for both Al-Mg-Y and Al-Fe-Y. Figure 8 presents two sets of bright field TEM micrographs for each $T_{HP}$ value, with the top row showing the location of these nanorods (indicated by red arrows) and the bottom row showing higher magnification views so that the dimension of these nanorods can be clearly seen. At $T_{HP}$ = 449 °C (Figure 8(a)), no nanorods are observed either in the grain interior or at grain boundaries. We note that the FCC grain size is approximately 20 nm, confirming the XRD result in Figure 2. As $T_{HP}$ is raised to 495 °C (Figure 8(b)), nanorods begin to form at grain boundaries with a length of ~10-20 nm and width of a few nanometers. When $T_{HP}$ is further increased to 540 °C, the size of these nanorods increases to a length of 20-30 nm and width of ~3-5 nm. The nanorods begin to form at grain boundaries and are first observed in these regions, but are not restricted by the adjacent grains. This can be seen in Figure 8(c), where the nanorod enclosed by the yellow square grows into an adjacent grain in the magnified view. This behavior is very different from amorphous complexions that are restricted along grain boundaries, and a detailed comparison between the nanorods and amorphous complexions will be discussed below. Finally, at the highest $T_{HP}$ of 585 °C (Figure 8(d)), the nanorods grow to ~50 nm in length, while the width stays below 10 nm. Therefore, with increasing $T_{HP}$, the size of these nanorods increases. As the nanorods grow, more and more of them extend into the grain interior. It should be noted that since no preferred crystallographic textures exist in the present alloy systems, these nanorods most likely orient randomly. In addition, the rod-like shape is the only one observed for dozens of features investigated across the multiple alloys and samples.

High resolution TEM was next used to investigate the internal structure of the nanorods. Figure 9(a) shows a high resolution TEM micrograph of one nanorod located at a grain boundary in Al-Ni-Y with $T_{HP}$ = 585 °C. The different lattice fringes on each side of the nanorod indicate that these regions correspond to two different grains and therefore this nanorod is located along a single grain boundary. Fast Fourier transform (FFT) diffractograms were performed on three enclosed areas in Figure 9(a), with two of the areas being within the adjacent grains while the other area including both one grain interior and the nanorod. We note that the width of the nanorod is too small in this case to only select that region, as selection of a smaller region results in an FFT diffractogram with poor resolution. The FFT diffractograms corresponding to the grain interior (Figures 9(b) and (d)) show different diffraction patterns, indicating two grains with different orientations, whereas no extra diffraction spots are observed within Figure 9(c) relative to that from Figure 9(d), suggesting that the nanorod is amorphous. To clarify whether the nanorods formed during the naturally cooled processing route or already existed from the high temperature consolidation, the annealed and quenched Al-Ni-Y pellet is examined, with a representative bright field TEM micrograph shown in Figure 9(e). Similar to the naturally cooled sample, nanorods are observed at grain boundaries (indicated by red arrows). However, compared to the slowly cooled sample, the dimension of these nanorods is much larger with lengths of ~50-100 nm and widths of ~10 nm (compared to lengths of 20-30 nm and widths of ~3-5 nm in the naturally cooled sample described above). Figure 9(f) is a high resolution TEM micrograph of one nanorod in the quenched sample, where one area entirely within the nanorod interior is enclosed and its corresponding FFT diffractogram is presented in Figure 9(g). This FFT image provides clear evidence that the nanorod is amorphous, as no diffraction spots are observed. Therefore, in both the naturally cooled and quenched samples, nanorods form at grain boundaries, with the size in the quenched sample larger than that in the naturally cooled sample, suggesting that the time allowed during natural cooling shrinks these features. The nanorods share several similar characteristics with the amorphous grain boundary complexions (Figures 6 and 7). First, both nucleate at grain boundaries once they are chemically enriched owing to segregation of the alloying species. Second, both the nanorods and amorphous complexions have disordered structures. Finally, the dimensions of both are smaller in the naturally cooled state than in the quenched state, suggesting that both structures prefer a larger equilibrium size at higher temperature. However, there is also a major difference between the nanorods and amorphous complexions. Unlike the amorphous complexions, which remain strictly bounded by adjacent crystalline grains, some of the nanorods grow into the grain interior (Figure 8).

Since the various grain boundary structures observed here presumably are formed due to grain boundary segregation of dopant elements, the distribution of dopant atoms is next studied. STEM-EDS was performed on the three alloys with $T_{HP}$ = 540 °C, and the corresponding elemental mapping is presented in Figure 10. The HAADF images in the first column show that nanorods are abundant in each alloy, and that their size is similar for all three alloys. In addition, the edges of nanorods appear much brighter than both the nanorod interior and Al matrix, suggesting that the edge should be rich in heavier elements since HAADF intensity is proportional to atomic number. The mapping of Ga can again be used for visualization of the interface between the nanorods and matrix. The corresponding mapping reveals that Al atoms are depleted at the nanorod edge, while the concentration of Y atoms is significantly enriched. The distribution of the alkaline-earth/transition metal dopants (Mg, Fe, or Ni) does not appear to be closely aligned with the nanorods in a discernable way, as Fe and Ni atoms are uniformly distributed within these EDS maps while small clusters of Mg atoms are observed. The Mg clusters are consistent with clustering of O atoms (not shown). Nanoscale dopant aggregation has been observed previously in Al alloys. For example, Liddicoat et al. [68] employed atom probe tomography to investigate the microstructure responsible for outstanding mechanical properties of a 7075 Al alloy processed by high-pressure torsion. Dopant atoms (Zn, Mg, and Cu) were found to segregate to grain boundaries in the form of two types of characteristic morphologies – the first one is a point-like "nodal" arrangement, and the second one is one-dimensional "lineal" structure. The lineal structure described by these authors has a length and diameter of ~ 17.5 nm and 4.4 nm, respectively, which are comparable to the dimensions of the nanorods in the present study.

However, in contrast to Ref. [68], the interior of the nanorods in the present study shows enrichment of elements different from that on the edge, as Al and C elements clearly segregate into the nanorod interior. The C in the system likely comes from the addition of stearic acid during ball milling and can be considered as an unintentional impurity. Therefore, the nanorods have a core-shell structure with the core composed of Al and C and the edges being Y-enriched.

Due to the small dimensions and disordered structure, neither the amorphous complexions nor core-shell nanorods are detected by XRD scans. However, XRD measurements do show that secondary phases emerge for all alloys during hot pressing, with their weight fraction as a function of $T_{HP}$ is shown in Figure 11. For all alloy systems, the secondary phases for the lowest $T_{HP}$ are different from those at higher temperatures. For $T_{HP}$ = 449 °C, all three alloys contain a small amount of yttrium hydrides, while Al-Fe-Y and Al-Ni-Y also contain binary intermetallic phases comprised of Al and the transition metal dopant. The hydride phases are likely formed during ball milling, due to the stearic acid contamination. When $T_{HP}$ was raised to 495 °C and above, a single secondary phase dominates and grows in each sample, which is $Al_3Y$, $Al_{10}Fe_2Y$, or $Al_{19}Ni_5Y_3$ for the Al-Mg-Y, Al-Fe-Y, or Al-Ni-Y system, respectively. The XRD crystallite size of these intermetallic phases increases from a range of ~10-30 nm for $T_{HP}$ = 495 °C to ~30-70 nm for $T_{HP}$ = 585 °C, while the weight fraction for the highest $T_{HP}$ reaches 11.0%, 23.2%, and 24.3% for $Al_3Y$, $Al_{10}Fe_2Y$, $Al_{19}Ni_5Y_3$, respectively. The different intermetallic phase in each system is consistent with the pair-wise enthalpy of mixing listed in Table 2. Overall, Al-Mg-Y has the least driving force to incorporate Mg into the intermetallic phase since $\Delta H_{mix}^{Al-Y}$ [39] is much more negative than $\Delta H_{mix}^{Al-Mg}$ [39] and $\Delta H_{mix}^{Mg-Y}$ [39], while the pair-wise enthalpies between Al and the dopant elements are more comparable in the other two alloy systems. Moreover, the grain boundary segregation tendency of Mg is stronger than that of Fe and Ni (Figure 5), so fewer Mg atoms are available to potentially form intermetallic phases. Figure 11(b) presents TEM selected area diffraction patterns for the highest $T_{HP}$ (585 °C), where the existence of all secondary phases from XRD is verified.

The intermetallic grain sizes measured by XRD are on the order of tens of nanometers, comparable to the Al-rich FCC phase, but this technique can only be used to accurately measure crystal size for nanoscale features and it is possible that larger precipitates are present (where XRD peak broadening would be undetectable). To understand the overall size distribution and spatial distribution of these intermetallic phases, backscattered electron (BSE) imaging is employed. Figure 12 presents BSE micrographs corresponding to $T_{HP}$ from 495 °C to 585 °C, where a single intermetallic phase dominates in each alloy. For $T_{HP}$ = 495 °C, magnified BSE images are also shown for more clear visualization of the shape of intermetallic phases, which are particles consisting of multiple grains. Due to the relatively higher effective atomic number, the intermetallic phases appear brighter in the backscattered electron micrographs, while the darker area corresponds to the Al matrix. The morphology of the intermetallic phases varies for the different alloys. In Al-Mg-Y, the $Al_3Y$ phase are very fine dispersed particles with a diameter on the order of tens to a few hundred nanometers. This is consistent with the observation made by Foley et al. [69] that the average size range of $Al_3Y$ particles is 20-100 nm in Al-Y alloys produced by a rapid solidification processing technique. In Al-Fe-Y, the $Al_{10}Fe_2Y$ particles also exhibit an equiaxed shape but with a slightly larger size compared to $Al_3Y$ particles in Al-Mg-Y. In contrast, the $Al_{19}Ni_5Y_3$ particles in the Al-Ni-Y are needle-shaped. Li et al. [70] also reported the existence of a rod or plate-like metastable $Al_{19}Ni_5Y_3$ phase formed in an Al-Ni-Y alloy, showing that the $Al_{19}Ni_5Y_3$ phase can form either during quenching from the liquid state at a lower cooling rate or from crystallization of amorphous alloys. Since the highest $T_{HP}$ in the present study is still below the melting temperature, it is possible that some $Al_{19}Ni_5Y_3$ particles are a result of crystallization of amorphous regions, such as the amorphous complexions that our Al alloys host.

When $T_{HP}$ increases, some intermetallic particles grow with the largest diameter being approximately 5 µm. However, most of the intermetallic particles remain small with sizes of a few hundred nanometers, and the shape of the intermetallic particles does not change with increasing $T_{HP}$. Figure 13 shows cumulative distribution functions of the intermetallic particle size for all three alloys. For Al-Mg-Y, the whole distribution shifts to larger particle size with increasing $T_{HP}$, while the curve corresponding to Al-Fe-Y also moves to larger size regime but to a lesser extent. In contrast, for Al-Ni-Y, only the extreme large tail of the distribution shows change as $T_{HP}$ increases, indicating that only a few $Al_{19}Ni_5Y_3$ particles grow while most of the particles remain the same size. Therefore, the intermetallic particle size is more stable in Al-Ni-Y than in Al-Mg-Y and Al-Fe-Y. Even though the average sizes of these intermetallic particles are only a few hundred nanometers, they are much larger than the grain size of the Al rich FCC phase. For nanocrystalline alloys, a kinetic approach to stabilizing grain size is by pinning the grain boundaries using precipitates, which is often referred to as Zener pinning [71]. According to Zener pinning theory, the maximum diameter of pinning particles that are effective for stabilizing grains is $d = (3/4)D \cdot f$, where $D$ is the grain size and $f$ is the volume fraction of pinning particles [72]. Since the volume fraction of pinning particles cannot be larger than 1, particles with a size comparable and/or much smaller than the matrix grain size should be effective in limiting grain growth. For example, Praveen et al. [73] fabricated a nanocrystalline CoCrFeNi high entropy alloy by mechanical alloying and spark plasma sintering, which exhibited high thermal stability after annealing at 900°C for 600 h. One of the dominant factors contributing to the good thermal stability in Ref. [73] was the Zener pinning effect from fine Cr-rich oxide precipitates with a size of approximately 30 nm, much smaller than the FCC matrix grains with an average size of 130 nm. In contrast, in the present study, the intermetallic particles are much larger than the Al matrix grains, so it is unlikely that these intermetallic particles contribute significantly to stabilizing the nanosized grains.

### 3.4 Mechanical behavior of nanocrystalline Al with hierarchical reinforcements

A hierarchical microstructure (including amorphous complexions, nanoscale amorphous core-shell precipitates, and intermetallic particles with sizes of a few hundred nanometers) was observed within the bulk nanocrystalline alloys studied here, which we hypothesize will give rise to excellent mechanical properties. To test this, nanoindentation tests were employed and Table 3 lists the hardness values for all three bulk alloys consolidated at $T_{HP}$ = 585 °C and then naturally cooled to room temperature. The Al-Mg-Y system exhibits the highest hardness of 2.77 ± 0.12 GPa, while Al-Fe-Y and Al-Ni-Y show slightly lower but still high hardness values of 2.18 ± 0.15 GPa and 2.29 ± 0.16 GPa, respectively. Therefore, even though Al-Mg-Y has a slightly larger Al matrix grain size than the other two alloys, it exhibits the highest hardness. Several studies showed that the Hall-Petch relation is not applicable to nanocrystalline materials with grain sizes smaller than ~10-20 nm because the grains are too small to support conventional intragranular dislocation plasticity [74]. However, the average grain size of the pellets with $T_{HP}$ = 585 °C is approximately 50 nm, which falls into the regime where the traditional Hall-Petch effect would still be expected to be operative. This suggests that other mechanisms in addition to grain size strengthening play an important role in increasing the strength of Al-Mg-Y. Xue et al. [75] observed a high density of growth twins in nanocrystalline Al-Mg thin films fabricated by DC magnetron sputtering, reporting that the strengthening effect from incoherent twin boundaries was equivalent to that from high-angle grain boundaries. Therefore, the higher hardness for Al-Mg-Y could possibly come from nanoscale twin boundaries. While we did not observe nanotwins in our microstructures, it is difficult to unequivocally rule out their existence in limited quantities. In addition, the different intermetallic phase in each system may also contribute to the various hardness values.

A comparison of hardness values of Al alloys from both the present work and other studies from the literature is shown in Figure 14. In general, all three of our alloys fabricated via a simple powder metallurgy approach exhibit hardness comparable to or higher than the other Al alloys, which range from nanocrystalline to micron-sized grains synthesized using different techniques. In fact, the Al-Mg-Y alloy in this work is the absolute hardest at 2.77 GPa. Youssef et al. [76] employed an in-situ mechanical alloying consolidation technique to fabricate a single-phase nanocrystalline Al-5 at.% Mg with a reported average grain size of 26 nm and hardness of 2.30 ± 0.19 GPa. That hardness value is four times higher than a conventional polycrystalline Al 5083 alloy with an average grain size of 5.5 μm and hardness of 0.57 GPa, which was mainly attributed to grain size refinement in Ref. [76]. Compared to the present ternary Al-Mg-Y with $T_{HP}$ = 585 °C, the hardness of the binary Al-Mg is significantly lower even though that material had a smaller grain size (as no high temperature was involved during the consolidation process).

The improved strength of the Al-Mg-Y presented in this study demonstrates the advantage of adding a rare earth element (or increasing the chemical complexity in general) to the nanocrystalline alloy that leads to the formation of a hierarchically reinforced microstructure. Rajulapati et al. [77] synthesized nanocrystalline Al-W alloys containing up to 4 at.% W using ball milling and then hot compaction at 573 K. The resulted alloy contained two phases: an Al phase and a W phase, both with average grain sizes of 34 ± 2 nm. The individual crystallites of the W phase were small in this material, but the W particle size (i.e., a region of material containing one or more aggregated W grains) ranged from a few tens of nanometers up to larger than 500 nm. The hardness of the Al-W alloy increased from 0.94 GPa with no W addition up to 1.23 GPa with 4 vol.% W, which was mainly attributed to Orowan hardening due to the smaller W particles in the system. In the present study, the size of the core-shell nanorods is smaller than the Al matrix grain size, and therefore may also have a Orowan hardening effect on the strength. In addition, in Ref. [68], the clusters of dopant atoms (Zn, Mg, and Cu) at the grain boundary region were suggested to play an important role in strengthening of a 7075 Al alloy, with the enriched regions having similar dimensions as the present amorphous core-shell nanorods, pointing to a strengthening effect of the nanorods. Moreover, in the present study, nanoindentation tests were also performed on the annealed and quenched Al-Ni-Y sample, which exhibits an average grain size of ~100 nm. The hardness is very close to the measured value for the naturally cooled counterpart, even though the mean grain size of the annealed sample is twice as large, indicating that grain size strengthening is not the dominant strengthening mechanism. As such, it is likely that the primary strengthening effect comes from the nanorod precipitates. In order to design solid solution strengthened Al alloys, Hung et al. [78] performed first-principle calculations to identify Al-Ce and Al-Co as promising candidates, and then synthesized the corresponding alloys using arc melting and produced non-equilibrium microstructures through laser surface glazing. The resultant microstructure included micron-sized matrix grains with nanoscale secondary phases, which exhibited hardness values of 1.26 GPa and 1.74 GPa for Al-1 at.% Ce and Al-1 at.% Co, respectively, comparable to a value of 1.505 GPa for commercial Al 6061-T6 alloys. In another study on ternary Al alloys, Aboulkhair et al. [79] performed nanoindentation to investigate the hardness of selective laser melted AlSi10Mg with micron-sized grains and obtained a value of 2.2 ± 0.1 GPa. This high value for the coarse-grained Al alloy was attributed to a homogeneous elemental distribution and finely dispersed Si particles with sizes of a few micrometers in the Al matrix, as opposed to Si flakes with a length of approximately a hundred micrometers and width of tens of micrometers in an as-cast samples which exhibited a lower hardness. For the alloys studied here, the high hardness is attributed to a nanocrystalline grain size and a hierarchical reinforcement microstructure consisting of nanometer thick amorphous complexions, nanoscale amorphous core-shell precipitates, and intermetallic particles with average sizes of a few hundred nanometers. By combining both grain size and obstacle strengthening, the combined effect is alloys which are harder than previously reported examples (see Figure 14). Moreover, the total amount of dopant atoms used here is only 4 at.% for each alloy, considerably lower than that (>10 at.%) in Ref. [79], thereby leading to alloys with exceptional strength and thermal stability while preserving the low densities required for advanced light-weight Al alloys.

## 4. Conclusions

In the present study, three ternary alloys (Al-Mg-Y, Al-Fe-Y, and Al-Ni-Y), were successfully synthesized into bulk samples with both high density (>99%) and nanocrystalline grains (<60 nm) with a hot pressing temperature of 92% of the melting temperature of pure Al. This was accomplished by choosing appropriate dopant species that can lead to both grain boundary segregation and amorphous complexion formation. The microstructures corresponding to four hot pressing temperatures of each alloy were examined in detail using XRD and TEM, while mechanical properties were tested using nanoindentation. The following conclusions are drawn:

1) All three alloy systems exhibit a superior combination of small grain size, full density, and bulk size that outperform previous reports, suggesting the importance of the grain boundary regions on alloy selection. The addition of a combination of an alkaline-earth/transition metal element (Mg, Fe, or Ni) and a rare-earth element (Y) significantly increases the atomic size mismatch between the three constituent elements, promoting grain boundary segregation and amorphous complexion formation that enable thermal stability of grain size and activated sintering. Moreover, the three alkaline-earth/transition metal elements show different segregation behavior and temperature ranges of activated sintering, suggesting the important role of these metals in controlling grain boundary activities during consolidation.

2) A hierarchical microstructure was observed in all systems, consisting of nanometer-thick amorphous grain boundary complexions, amorphous core-shell nanorods with Y-enriched edges and Al+C-enriched interiors, and intermetallic particles with average sizes of a few hundred nanometers. The core-shell nanorods are observed to nucleate at grain boundaries, which share similarities with amorphous complexions but also possess their own unique properties, such as not being confined by adjacent grains.

3) The retention of the amorphous grain boundary complexions, even with a very slow cooling rate, shows that these features are much more resistant to transitioning back to the ordered grain boundary complexions stable at room temperature than amorphous complexions observed in other systems such as Cu-Zr, Cu-Hf, and Cu-Zr-Hf.

4) The hierarchical microstructure gives rise to exceptional hardness exceeding other reports for bulk Al alloys without sacrificing the light-weight property, since only a total of 4 at.% dopant atoms were added.

The results of this study generally shed light on a design pathway to fabricate large-scale nanocrystalline Al alloys strengthened by hierarchical reinforcements, which have great potential for structural applications. Grain size must be stabilized while densification encouraged, all while seeding the precipitation of reinforcing phases with multiple length scales. Grain boundary segregation can serve as a template for all of these behaviors.

**Declaration of Competing Interest**

The authors declare that they have no known competing financial interest or personal relationships that could have appeared to influence the work reported in this paper.

**Acknowledgements**

This work was supported by the U.S. Department of Energy, Office of Energy Efficiency and Renewable Energy (EERE), under the Advanced Manufacturing Office Award No. DE-EE0009114. The authors acknowledge the use of facilities and instrumentation at the UC Irvine Materials Research Institute (IMRI), which is supported in part by the National Science Foundation through the UC Irvine Materials Research Science and Engineering Center (DMR-2011967). SEM, FIB, and EDS work was performed using instrumentation funded in part by the National Science Foundation Center for Chemistry at the Space-Time Limit (CHE-0802913).

# References

[1] G. S. Cole, A. M. Sherman, Light weight materials for automotive applications, Mater. Charact. 31(1), 3-9 (1995).

[2] Y. H. Zhao, X. Z. Liao, Z. Lin, R. Z. Valiev, and Y. T. Zhu, Microstructures and mechanical properties of ultrafine grained 7075 Al alloy processed by ECAP and their evolutions during annealing, Acta Mater. 52(15), 4589-4599 (2004).

[3] J. C. Williams, E. A. Starke, Jr., Progress in structural materials for aerospace systems, Acta Mater. 51, 5775-5799 (2003).

[4] C. B. Fuller, M. W. Mahoney, M. Calabrese, and L. Micona, Evolution of microstructure and mechanical properties in naturally aged 7050 and 7075 Al friction stir welds, Mater. Sci. Eng. A 527, 2233-2240 (2010).

[5] Q. Shi, Y. Huo, T. Berman, B. Ghaffari, M. Li, and J. Allison, Distribution of transition metal elements in an Al-Si-Cu-based alloy, Scr. Mater. 190, 97-102 (2021).

[6] K. Ma, H. Wen, T. Hu, T. D. Topping, D. Isheim, D. N. Seidman, E. J. Lavernia, and J. M. Schoenung, Mechanical behavior and strengthening mechanisms in ultrafine grain precipitation-strengthened aluminum alloy, Acta Mater. 62, 141-155 (2014).

[7] Q. Li, S. Xue, Y. Zhang, X. Sun, H. Wang, and X. Zhang, Plastic anisotropy and tension-compression asymmetry in nanotwinned Al-Fe alloys: An in-situ micromechanical investigation, Int. J. Plast. 132, 102760 (2020).

[8] B. Ahn, A. P. Newbery, E. J. Lavernia, and S. R. Nutt, Effect of degassing temperature on the microstructure of a nanocrystalline Al-Mg alloy, Mater. Sci. Eng. A 463, 61-66 (2007).

[9] H. A. Padilla II and B. L. Boyce, A review of fatigue behavior in nanocrystalline metals, Exp. Mech. 50, 5-23 (2010).

[10] A. Devaraj, W. Wang, R. Vemuri, L. Kovarik, X. Jiang, M. Bowden, J. R. Trelewicz, S. Mathaudhu, and A. Rohatgi, Grain boundary segregation and intermetallic precipitation in coarsening resistant nanocrystalline aluminum alloys, Acta Mater. 165, 698-708 (2019).

[11] J. Weissmuller, Alloy effects in nanostructures, Nanostructured Materials 3, 261-272 (1993).

[12] P. C. Millett, R. P. Selvam, and A. Saxena, Stabilizing nanocrystalline materials with dopants, Acta Mater. 55(7), 2329-2336 (2007).

[13] P. Choi, M. da Silva, U. Klement, T. Al-Kassab, and R. Kirchheim, Thermal stability of electrodeposited nanocrystalline Co-1.1 at.% P, Acta Mater. 53(16), 4473-4481 (2005).

[14] T. Chookajorn, H. A. Murdoch, and C. A. Schuh, Design of Stable Nanocrystalline Alloys, Science 337, 951-954 (2012).

[15] H. A. Murdoch and C. A. Schuh, Estimation of grain boundary segregation enthalpy and its role in stable nanocrystalline alloy design, J. Mater. Res. 28(16), 2154-2163 (2013).

[16] S. G. Kim and Y. B. Park, Grain boundary segregation, solute drag and abnormal grain growth, Acta Mater. 56, 3739-3753 (2008).

[17] G. H. Balbus, J. Kappacher, D. J. Sprouster, F. Wang, J. Shin, Y. M. Eggeler, T. J. Rupert, J. R. Trelewicz, D. Kiener, V. Maier-Kiener, D. S. Gianola, Disordered interfaces enable high temperature thermal stability and strength in a nanocrystalline aluminum alloy, Acta Mater. 215, 116973 (2021).

[18] Y. Hu and T. J. Rupert, Atomistic modeling of interfacial segregation and structural transitions in ternary alloys, J. Mater. Sci. 54, 3975-3993 (2019).

[19] W. Xing, A. R. Kalidindi, D. Amram, and C. A. Schuh, Solute interaction effects on grain boundary segregation in ternary alloys, Acta Mater. 161, 285-294 (2018).

[20] W. Xing, S. A. Kube, A. R. Kalidindi, D. Amram, J. Schroers, and C. A. Schuh, Stability of ternary nanocrystalline alloys in the Pt-Pd-Au system, Materialia 8, 100449 (2019).

[21] P. R. Cantwell, T. Frolov, T. J. Rupert, A. R. Krause, C. J. Marvel, G. S. Rohrer, J. M. Rickman, and M. P. Harmer, Grain boundary complexion transitions, Annu. Rev. Mater. Res. 50, 465-492 (2020).

[22] P. R. Cantwell, M. Tang, S. J. Dillon, J. Luo, G. S. Rohrer, M. P. Harmer, Grain boundary complexions, Acta Mater. 62, 1-48 (2014).

[23] C. M. Grigorian and T. J. Rupert, Thick amorphous complexion formation and extreme thermal stability in ternary nanocrystalline Cu-Zr-Hf alloys, Acta Mater. 179, 172-182 (2019).

[24] J. Luo, Liquid-like interface complexion: From activated sintering to grain boundary diagrams, Curr. Opin. Solid State Mater. Sci. 12, 81-88 (2008).

[25] V. K. Gupta, D. -H. Yoon, H. M. Meyer III, and J. Luo, Thin intergranular films and solid-state activated sintering in nickel-doped tungsten, Acta Mater. 55, 3131-3142 (2007).

[26] O. K. Donaldson and T. J. Rupert, Amorphous Intergranular Films Enable the Creation of Bulk Nanocrystalline Cu-Zr with Full Density, Adv. Eng. Mater. 21(9), 1900333 (2019).

[27] C. M. Grigorian and T. J. Rupert, Critical cooling rates for amorphous-to-ordered complexion transitions in Cu-rich nanocrystalline alloys, Acta Mater. 206, 116650 (2021).

[28] J. D. Schuler and T. J. Rupert, Materials selection rules for amorphous complexion formation in binary metallic alloys, Acta Mater. 140, 196-205 (2017).

[29] A. Inoue, Stabilization of metallic supercooled liquid and bulk amorphous alloys, Acta Mater. 48, 279-306 (2000).

[30] T. Egami and Y. Waseda, Atomic size effect on the formability of metallic glasses, J. Non-Cryst. Solids 64, 113-134 (1984).

[31] Z. P. Lu, C. T. Liu, and W. D. Porter, Role of yttrium in glass formation of Fe-based bulk metallic glasses, Appl. Phys. Lett. 83(13), 2581-2583 (2003).

[32] H. Okamoto, Al-Mg (aluminum-magnesium), J. Phase Equilibria Diffus. 19(6), 598 (1998).

[33] K. Han, I. Ohnuma, and R. Kainuma, Experimental determination of phase equilibria of Al-rich portion in the Al-Fe binary system, J. Alloys Compd. 668, 97-106 (2016).

[34] H. Okamoto, Al-Ni (aluminum-nickel), J. Phase Equilibria Diffus. 25(4), 394 (2004).

[35] S. Liu, Y. Du, H. Xu, C. He, and J. C. Schuster, Experimental investigation of the Al-Y phase diagram, J. Alloys Compd. 414, 60-65 (2006).

[36] G. H. Balbus, F. Wang, and D. S. Gianola, Suppression of shear localization in nanocrystalline Al-Ni-Ce via segregation engineering, Acta Mater. 188, 63-78 (2020).

[37] S. C. Pun, W. Wang, A. Khalajhedayati, J. D. Schuler, J. R. Trelewicz, and T. J. Rupert, Nanocrystalline Al-Mg with extreme strength due to grain boundary doping, Mater. Sci. Eng. A 696, 400-406 (2017).

[38] T. T. Sasaki, T. Mukai, and K. Hono, A high-strength bulk nanocrystalline Al-Fe alloy processed by mechanical alloying and spark plasma sintering, Scr. Mater. 57(3), 189-192 (2007).

[39] A. Takeuchi and A. Inoue, Calculations of mixing enthalpy and mismatch entropy for ternary amorphous alloys, Mater. Trans. JIM 41(11), 1372-1378 (2011).

[40] J. M. Park, J. H. Na, D. H. Kim, K. B. Kim, N. Mattern, U. Kühn, and J. Eckert, Medium range ordering and its effect on plasticity of Fe–Mn–B–Y–Nb bulk metallic glass, Philos. Mag. 90(19), 2619-2633 (2010).

[41] Lin, Chun-Ming, and Hsien-Lung Tsai. Evolution of microstructure, hardness, and corrosion properties of high-entropy Al0.5CoCrFeNi alloy, Intermetallics 19(3), 288-294 (2011).

[42] R. Brandt and G. Neuer, Electrical Resistivity and Thermal Conductivity of Pure Aluminum and Aluminum Alloys up to and above the Melting Temperature, Int. J. Thermophys. 28(5), 1429-1446.

[43] D. Nath, F. Singh, and R. Das, X-ray diffraction analysis by Williamson-Hall, Halder-Wagner and size-strain plot methods of CdSe nanoparticles – a comparative study, Mater. Chem. Phys. 239, 122021 (2020).

[44] R. A. Young and D. B. Wiles, Profile shape functions in Rietveld refinements, J. Appl. Crystallogr. 15, 430-438 (1982).

[45] ASTM, Standard Practice for Determining the Inclusion or Second-Phase Constituent Content of Metal by Automatic Image Analysis, E 1245-03, 2003

[46] C. A. Schneider, W. S. Rasband, and K. W. Eliceiri, NIH Image to ImageJ: 25 years of image analysis, Nat. Methods 9, 671-675 (2012).

[47] L. A. Giannuzzi, J. L. Drown, S. R. Brown, R. B. Irwin, and F. A. Stevie, Applications of the FIB lift-out technique for TEM specimen preparation Microsc. Res. Tech. 41, 285-290 (1998)

[48] Agilent Technologies Nano Indenter G200 User's Guide.

[49] W. Hana, K. Maung, M. Enayati, J. C. Earthman, and F. A. Mohamed, Grain size stability in a cryomilled nanocrystalline Al alloy powders containing diamantane nanoparticles, Mater. Sci. Eng. A 746, 290-299 (2019).

[50] J. Luo and X. Shi, Grain boundary disordering in binary alloys, Appl. Phys. Lett. 92, 101901 (2008)

[51] X. Shi and J. Luo, Grain boundary wetting and prewetting in Ni-doped Mo, Appl. Phys. Lett. 94, 251908 (2009).

[52] J. Nie, J. M. Chan, M. Qin, N. Zhou, and J. Luo, Liquid-like grain boundary complexion and sub-eutectic activated sintering in CuO-doped $TiO_2$, Acta Mater. 130, 329-338 (2017).

[53] T. T. Ssaki, T. Ohkubo, and K. Hono, Microstructure and mechanical properties of bulk nanocrystalline Al-Fe alloy processed by mechanical alloying and spark plasma sintering, Acta Mater. 57, 3529-3538 (2009).

[54] G. M. Le, A. Godfrey, N. Hansen, W. Liu, G. Winther, and X. Huang, Influence of grain size in the near-micrometre regime on the deformation microstructure in aluminum, Acta Mater. 61, 7072-7086 (2013).

[55] M. Furukawa, Z. Horita, M. Nemoto, R. Z. Valiev, and T. G. Langdon, Microhardness measurements and the hall-petch relationship in an Al-Mg alloy with submicrometer grain size, Acta Mater. 44, 4619-4629 (1996).

[56] J. A. Haber and W. E. Buhro, Kinetic Instability of Nanocrystalline Aluminum Prepared by Chemical Synthesis; Facile Room-Temperature Grain Growth, J. Am. Chem. Soc. 120, 10847-10855 (1998).

[57] T. D. Topping, B. Ahn, Y. Li, S. R. Nutt, and E. J. Lavernia, Influence of process parameters on the mechanical behavior of an ultrafine-grained Al alloy, Metall. Mater. Trans. A 43A 505-519 (2012).

[58] K. A. Unocic, M. J. Mills, and G. S. Daehn, Effect of gallium focused ion beam milling on preparation of aluminum thin foils, J. Microsc. 240(3), 227-238 (2010).

[59] R. I. Babicheva, S. V. Dmitriev, Y. Zhang, S. W. Kok, N. Srikanth, B. Liu, and K. Zhou, Effect of grain boundary segregations of Fe, Co, Cu, Ti, Mg and Pb on small plastic deformation of nanocrystalline Al, Comput. Mater. Sci. 98, 410-416 (2015).

[60] L. Pauling, Atomic radii and interatomic distances in metals, J. Am. Chem. Soc. 69(3), 542-553 (1947).

[61] D. R. Clarke, On the detection of thin intergranular films by electron microscopy, Ultramicroscopy 4, 33-44 (1979).

[62] A. Khalajhedayati, Z. Pan, and T. J. Rupert, Manipulating the interfacial structure of nanomaterials to achieve a unique combination of strength and ductility, Nat. Commun. 7:10802, 1-8 (2016).

[63] D. Turnbull, Under what conditions can a glass be formed? Contemp. Phys. 10, 473-488 (1969).

[64] A. Peker and W. L. Johnson, A highly processable metallic glass: $Zr_{41.2}Ti_{13.8}Cu_{12.5}Ni_{10.0}Be_{22.5}$, Appl. Phys. Lett. 63, 2342 (1993).

[65] H. Men and D. H. Kim, Fabrication of ternary Mg-Cu-Gd bulk metallic glass with high glass-forming ability under air atmosphere, J. Mater. Res. 18, 1502-1504 (2003).

[66] J. Schroers and W. L. Johnson, Crystallization kinetics of the bulk-glass-forming Pd43Ni10Cu27P20 melt, Appl. Phys. Lett. 77, 1158 (2000).

[67] P. Garg, Z. Pan, V. Turlo, and T. J. Rupert, Segregation competition and complexion coexistence within a polycrystalline grain boundary network, Acta Mater. 218, 117213, 1-14 (2021).

[68] P. V. Liddicoat, X. -Z. Liao, Y. Zhao, Y. Zhu, M. Y. Murashkin, E. J. Lavernia, R. Z. Valiev, and S. P. Ringer, Nanostructural hierarchy increases the strength of aluminum alloys, Nat. Commun. 1:63 (2010).

[69] J. C. Foley, J. H. Perepezko, D. J. Skinner, Formation of metastable L12-Al3Y through rapid solidification processing, Mater. Sci. Eng. A 179/180 205-209 (1994)

[70] Y. Li, K. Georgarakis, S. Pang, F. Charlot, A. Lemoulec, S. Brice-Profeta, T. Zhang, and A. R. Yavari, Chill-zone aluminum alloys with GPa strength and good plasticity, J. Mater. Res. 24(4) 1513- 1521 (2009)

[71] C. C. Koch, R. O. Scattergood, M. Saber, and H. Kotan, High temperature stabilization of nanocrystalline grain size: Thermodynamic versus kinetic strategies, J. Mater. Res. 28(13), 1785-1791 (2013).

[72] C. S. Smith, Grains, phase, and interfaces: An introduction of microstructure. Trans. Metall. Soc. AIME 175, 15-51 (1948).

[73] S. Praveen, J. Basu, S. Kashyap, and R. S. Kottada, Exceptional resistance to grain growth in nanocrystalline CoCrFeNi high entropy alloy at high homologous temperatures, J. Alloys Compd. 662, 361-367 (2016).

[74] C. E. Carlton, and P. J. Ferreira, What is behind the inverse Hall-Petch effect in nanocrystalline materials? Acta Mater. 55, 3749-3756 (2007).

[75] S. Xue, Q. Li, Z. Fan, H. Wang, Y. Zhang, J. Ding, H. Wang, and X. Zhang, Strengthening mechanisms and deformability of nanotwinned AlMg alloys, J. Mater. Res. 33, 3739-3749 (2018).

[76] K. M. Youssef, R. O. Scattergood, K. L. Murty, and C. C. Koch, Nanocrystalline Al-Mg alloy with ultrahigh strength and good ductility, Scr. Mater. 54, 251-256 (2006).

[77] K. V. Rajulapati, R. O. Scattergood, K. L. Murty, Z. Horita, T. G. Langdon, and C. C. Koch, Mechanical properties of bulk nanocrystalline Aluminum-Tungsten Alloys, Metall. Mater. Trans. A 39A, 2528-2534 (2008).

[78] C. J. Hung, S. K. Nayak, Y. Sun, C. Fennessy, V. K. Vedula, S. Tulyani, S.-W. Lee, S. P. Alpay, and R. J. Hebert, Novel Al-X alloys with improved hardness, Mater. Des. 192, 108699 (2020).

[79] N. T. Aboulkhair, I. Maskery, C. Tuck, I. Ashcroft, and N. Everitt, Nano-hardness and microstructure of selective laser melted AlSi10Mg scan tracks, Industrial Laser Applications Symposium (ILAS 2015) 9657, 965702. International Society for Optics and Photonics (2015).

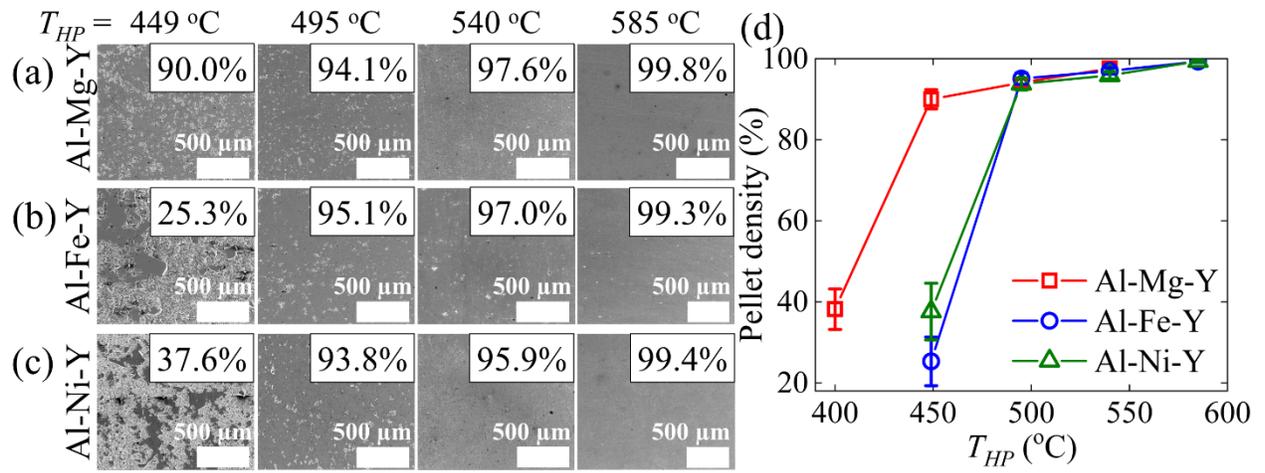

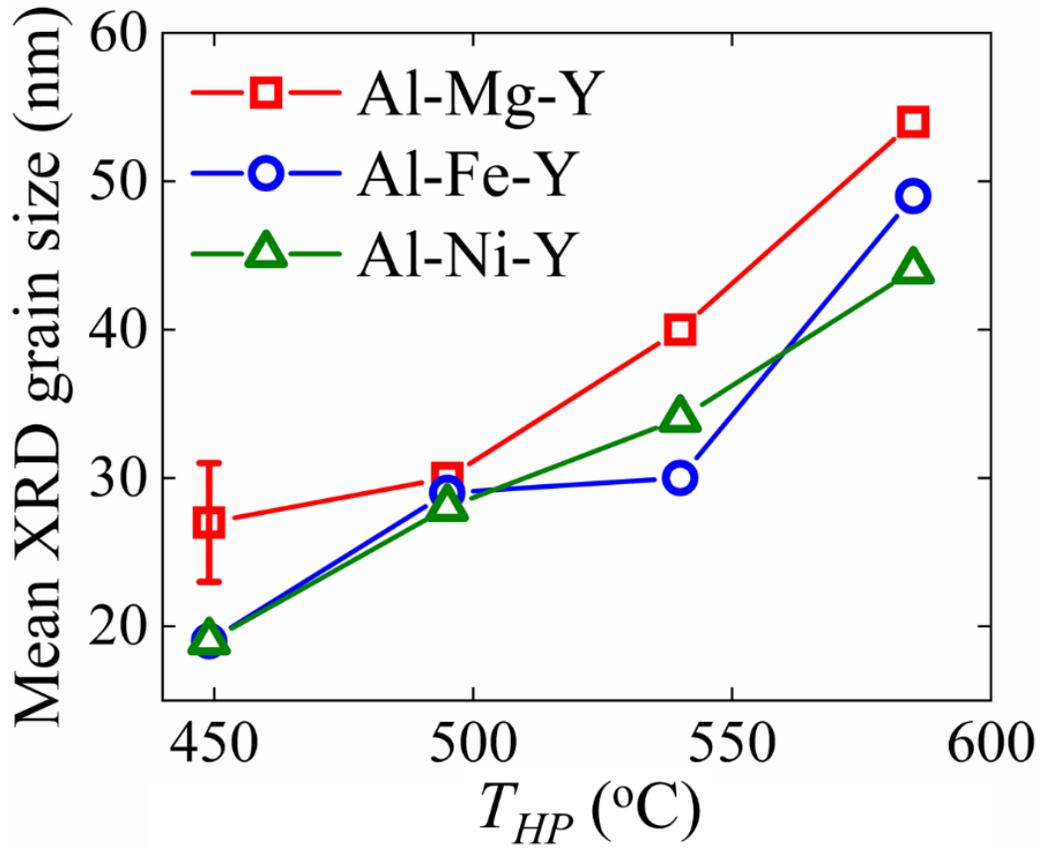

**Figure 2.** Mean grain size of the Al-rich FCC phase, as measured by XRD, as a function of $T_{HP}$ for Al-Mg-Y, Al-Fe-Y, and Al-Ni-Y pellets. Al-Mg-Y exhibits a slightly larger grain size than the other two alloys for all $T_{HP}$ values. However, all grain sizes remain well below 60 nm, even for the highest $T_{HP}$ (585 °C).

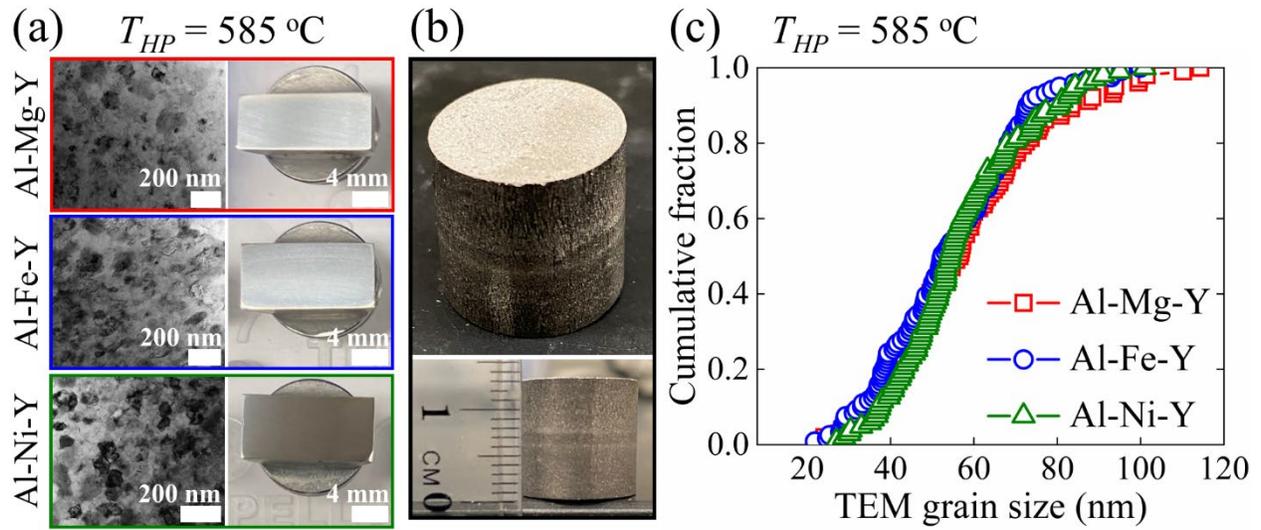

**Figure 3.** (a) Bright-field TEM micrographs and cross-sectional optical images of pellets for Al-Mg-Y, Al-Fe-Y, and Al-Ni-Y made with the highest $T_{HP}$ (585 °C). (b) Optical images of one as-consolidated pellet before polishing. (c) Cumulative distribution functions of grain sizes for the Al-rich FCC phase, as measured by TEM, for the highest $T_{HP}$. The average TEM grain sizes are measured to be 58 ± 19 nm, 54 ± 17 nm, and 57 ± 16 nm for Al-Mg-Y, Al-Fe-Y, and Al-Ni-Y, respectively, which are consistent with the XRD measurements.

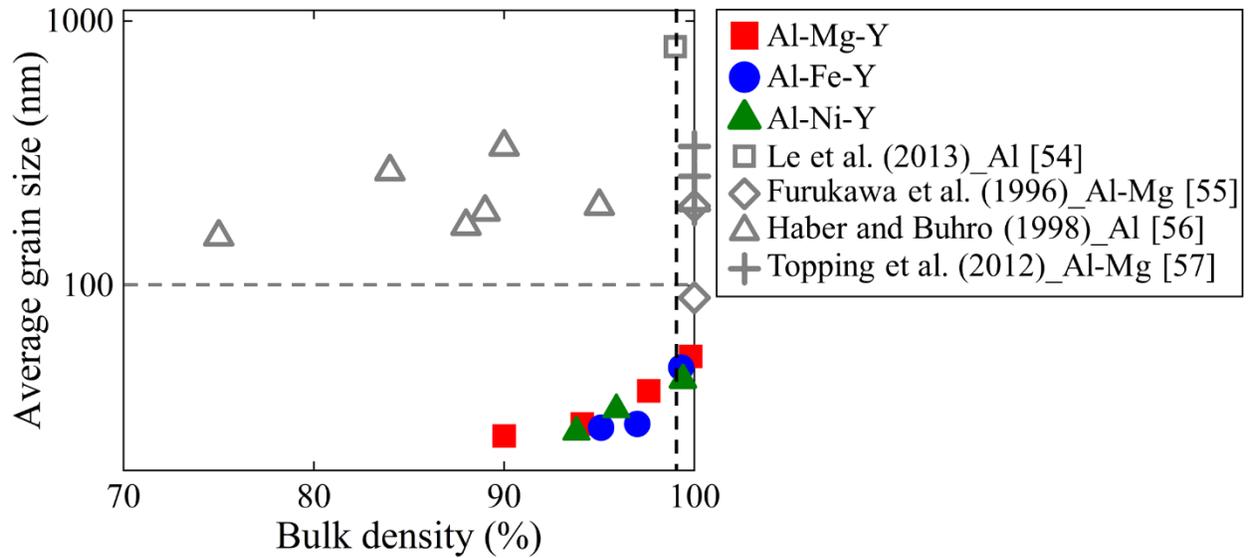

**Figure 4.** Comparison of TEM grain size versus bulk density between the three Al alloys in the present study, with Al and Al-rich alloys reported by other work included for reference. For the present Al alloys, only the samples after the appearance of activated sintering are shown. In general, all three alloys studied here show a superior combination of small grain size and high bulk density. The horizontal and vertical dashed lines mark a grain size of 100 nm and 99% pellet density, respectively.

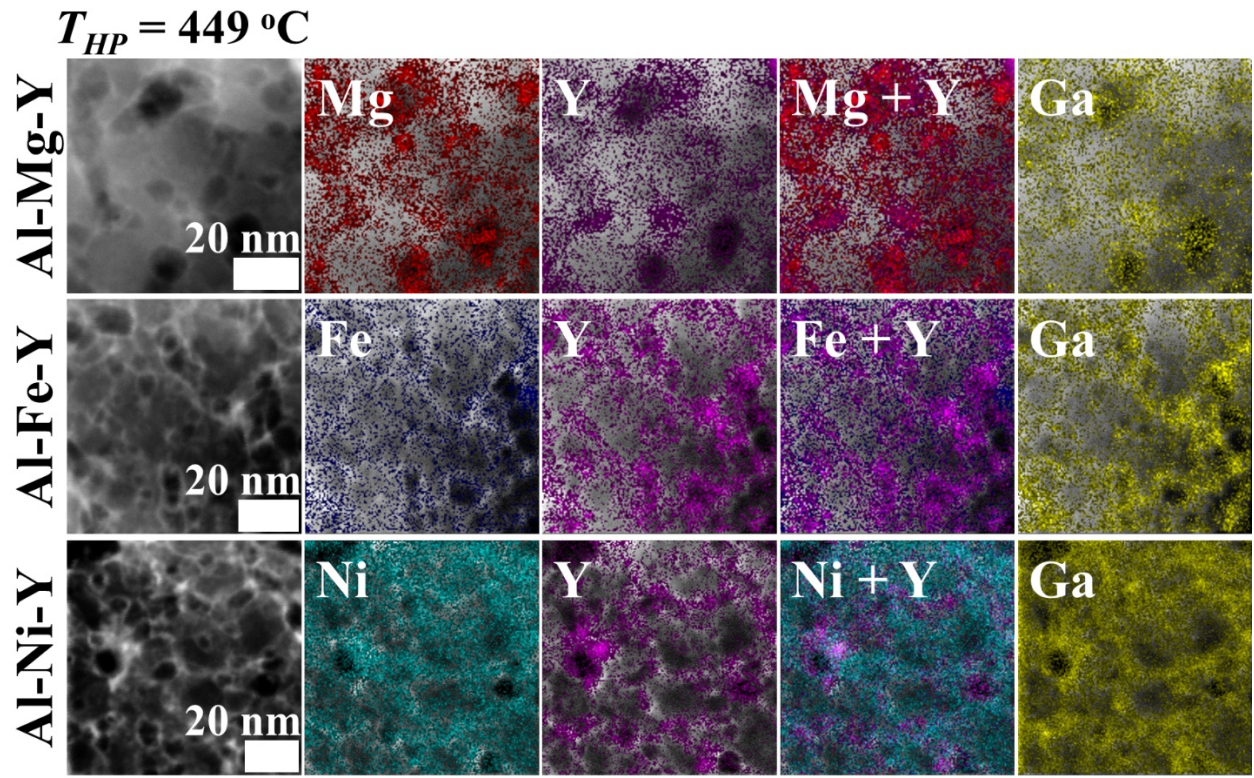

**Figure 5.** HAADF-STEM micrographs and the corresponding EDS elemental mapping of dopant atoms for Al-Mg-Y, Al-Fe-Y, and Al-Ni-Y pellets with the lowest $T_{HP}$ (449 °C). The distribution of all dopant atoms is consistent with that of Ga atoms, which are known to segregate to interfaces in Al and are therefore used for identification of grain boundary locations.

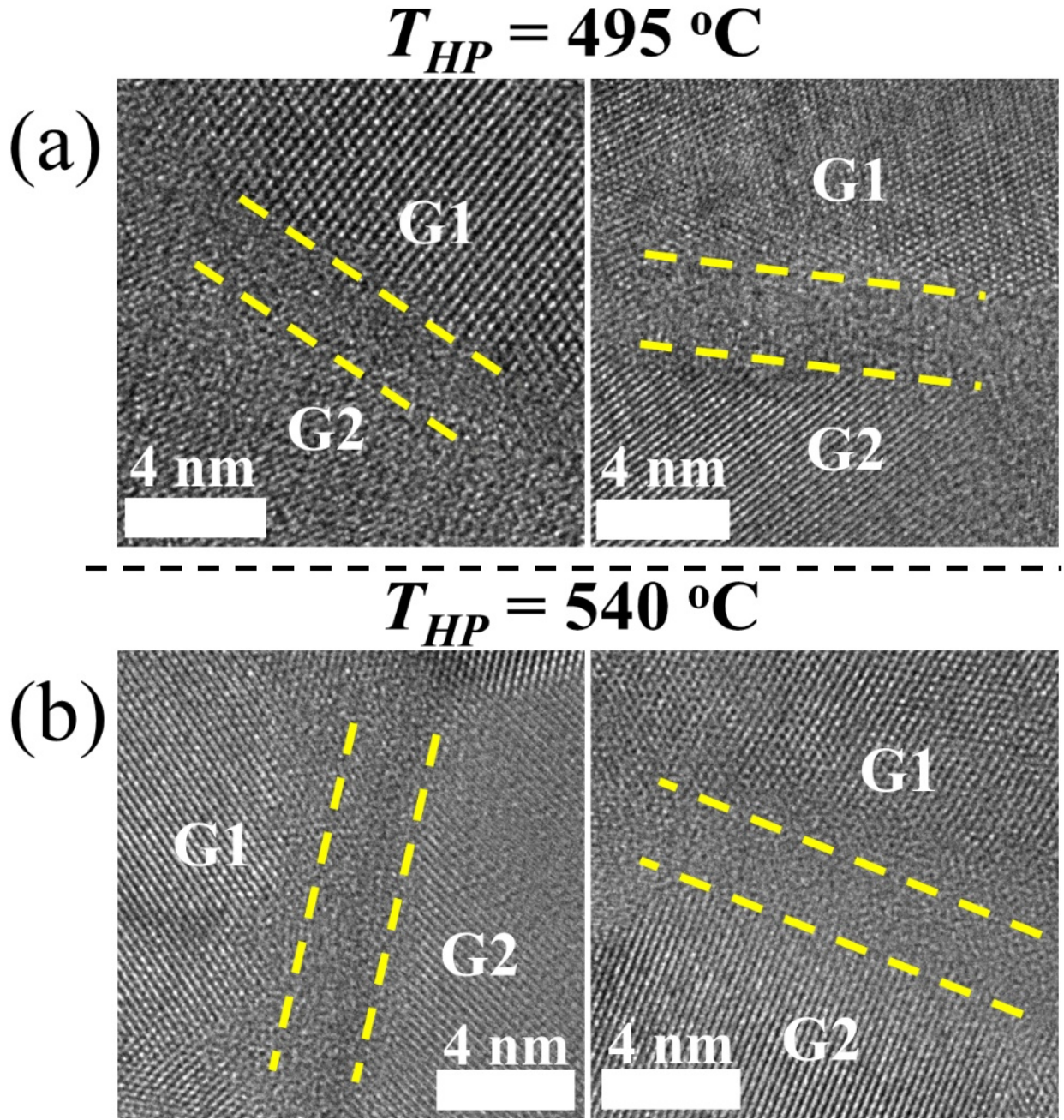

**Figure 6.** High resolution TEM micrographs of representative amorphous complexions in Al-Ni-Y pellets that were naturally cooled to room temperature after hot pressing. Two $T_{HP}$ values, (a) 495 °C and (b) 540 °C, above the observation of activated sintering are shown here. The amorphous complexions are outlined by yellow dashed lines while adjacent grains are denoted by "G1" and "G2".

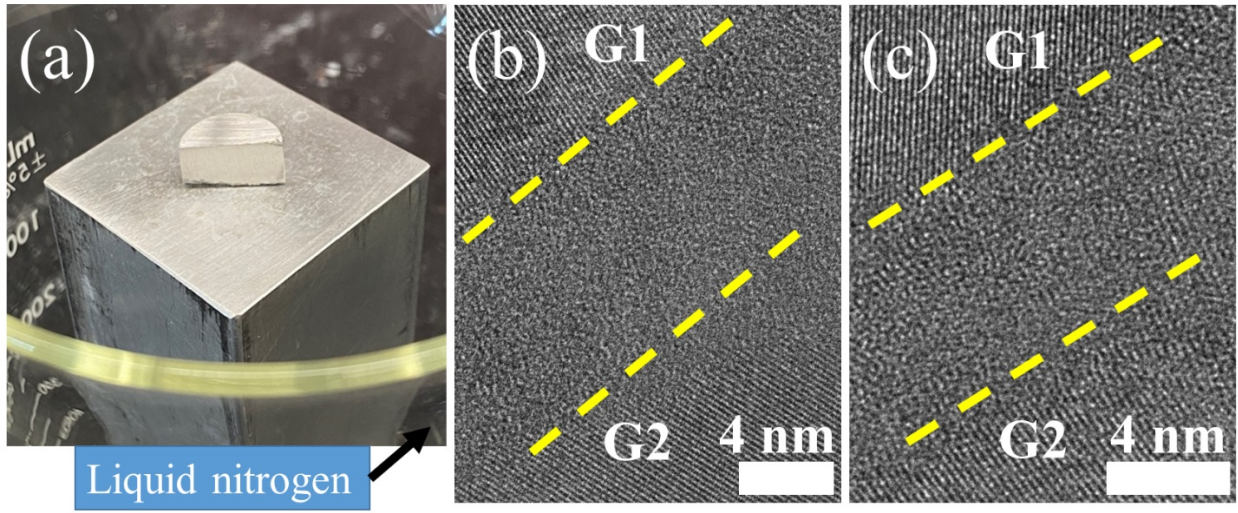

**Figure 7.** (a) Setup for quenching a pellet specimen. A pellet consolidated at $T_{HP}$ = 540 °C was subsequently annealed at 540 °C for 10 min, and then one half pellet was placed on an Al block submerged in liquid nitrogen as soon as possible so that high-temperature microstructures can be frozen in to be studied. (b) and (c) show high resolution TEM micrographs of two representative amorphous complexions observed in the quenched Al-Ni-Y pellet. The amorphous complexions are outlined by yellow dashed lines while adjacent grains are denoted by "G1" and "G2".

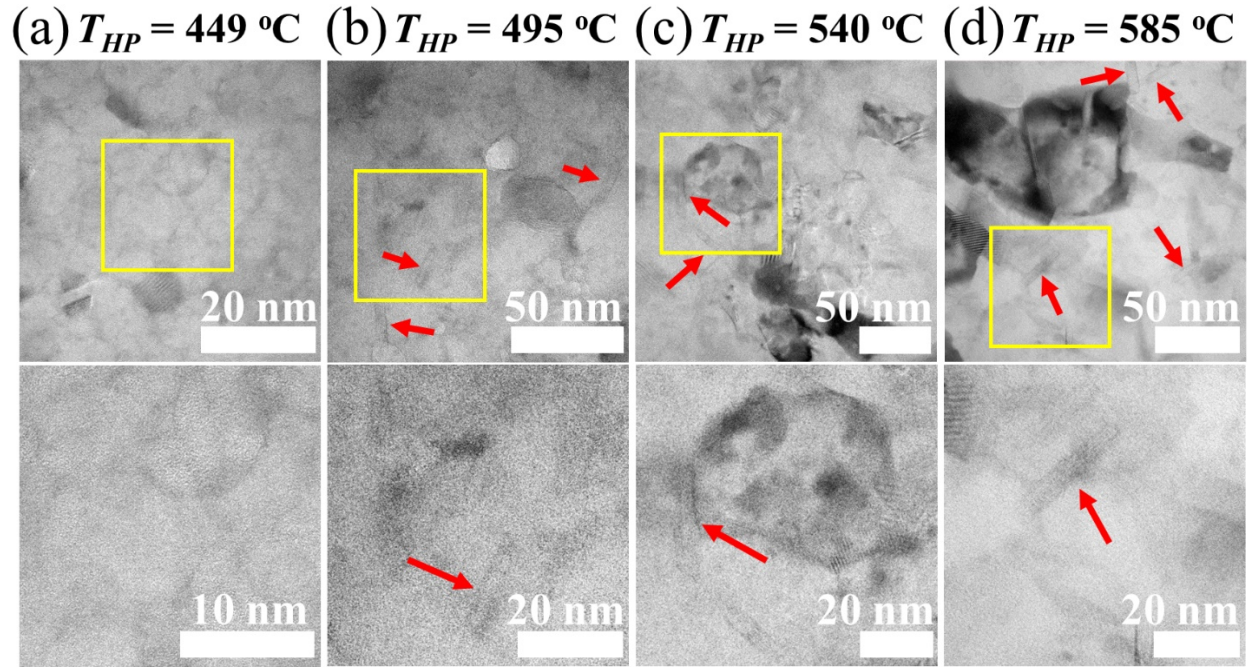

**Figure 8.** (a)-(d) Bright-field TEM micrographs and the corresponding magnified image showing the evolution of nanorods in naturally cooled Al-Ni-Y pellets with increasing $T_{HP}$. For $T_{HP} = 449$ °C, the microstructure only consists of nanocrystalline grains with a size of ~20 nm, consistent with the XRD measurement. For higher $T_{HP}$, nanorods (indicated by red arrows) begin to form at grain boundaries. The length of the nanorods increases from ~10-20 nm for $T_{HP} = 495$ °C to ~50 nm for $T_{HP} = 585$ °C.

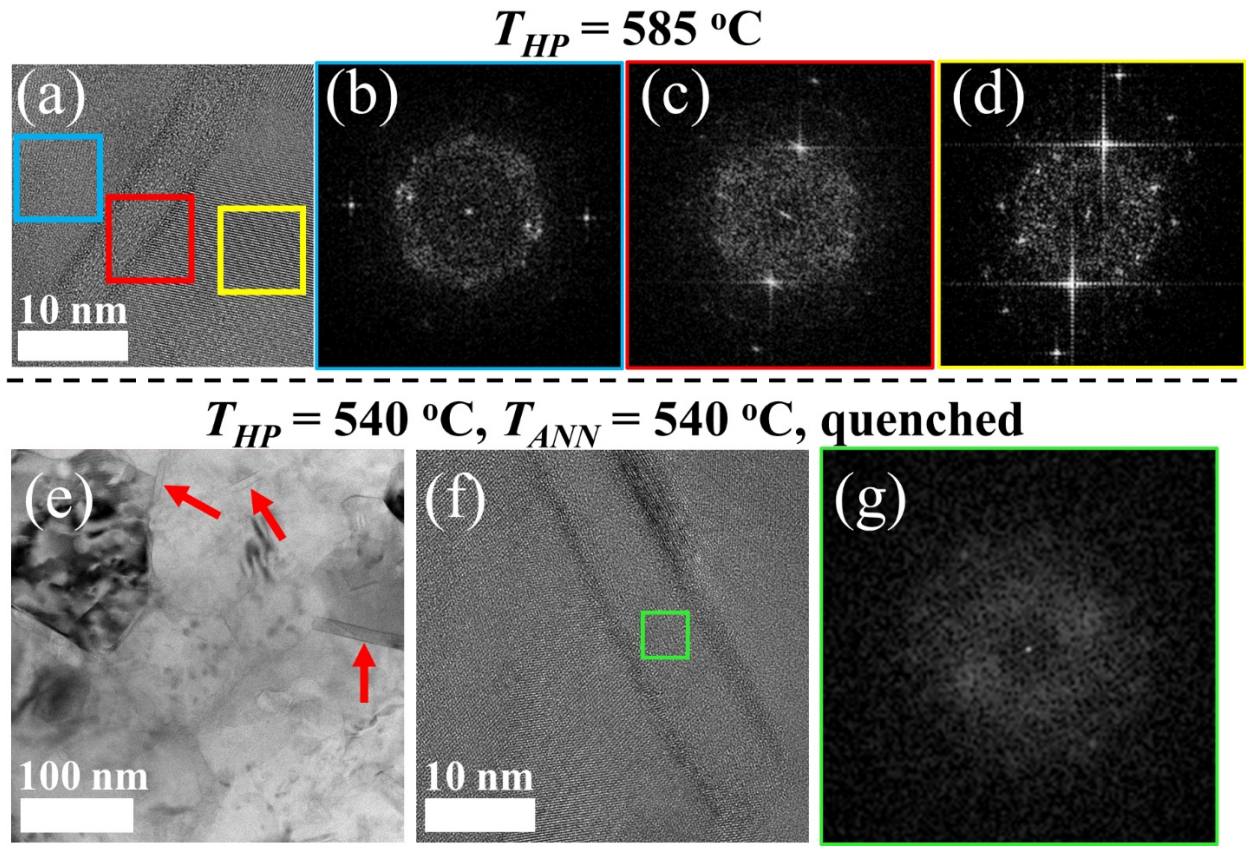

**Figure 9.** (a) High resolution TEM micrograph of one nanorod observed in a naturally cooled Al-Ni-Y pellet with the highest $T_{HP}$ (585 °C). Two areas in the adjacent grains and one area including both part of a grain and the nanorod interior are marked, with the corresponding FFTs shown in (b)-(d). The FFTs for the two grains in (b) and (d) show different diffraction spots, indicating that they are two different grains. No extra diffraction spots were observed in (c) compared to (d), suggesting that the nanorod has an amorphous structure. (e) Bright-field TEM micrograph taken from an annealed and quenched Al-Ni-Y pellet, where examples of nanorods are indicated by red arrows. The nanorods have a length of ~100 nm after annealing and quenching. (f) high resolution TEM image of one nanorod in the annealed and quenched sample with (e) an FFT taken from an area fully within the nanorod being enclosed. No diffraction spots are observed, confirming the disordered interior of the nanorods.
46

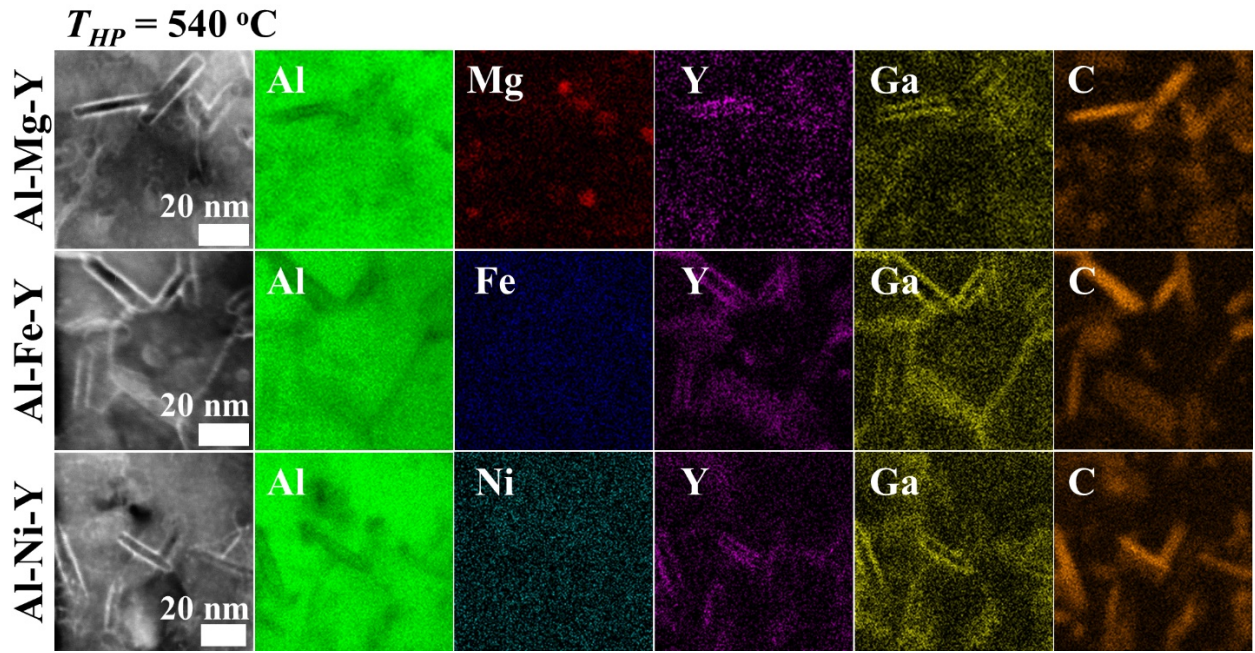

**Figure 10.** HAADF-STEM micrographs and the corresponding EDS elemental mapping for $T_{HP}$ = 540 °C showing the chemistry of the nanorods in all three alloys. The edge of the nanorods appears much brighter than the nanorod interiors and matrix, suggesting enrichment in heavier-elements. The map of Ga atoms is provided to help identify interfaces, which in this case are primarily the edges of the nanorods. In all three alloys, the concentration of Y atoms is consistent with the map of Ga atoms, while Fe and Ni are uniformly distributed in the whole area. Some Mg atoms form small clusters, which are consistent with the clustering of O atoms (not shown). Al and C atoms are found to primarily constitute the nanorod interiors.

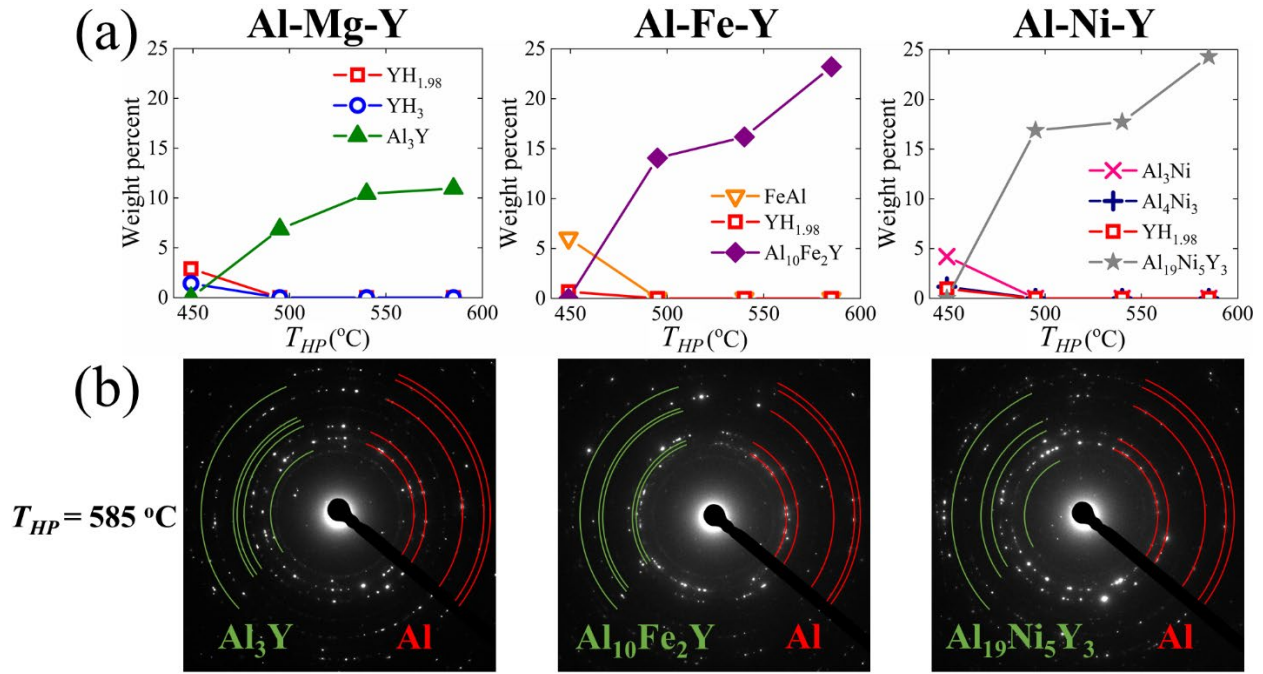

**Figure 11.** (a) Weight fraction of secondary phases in each alloy, as measured from XRD and presented as a function of $T_{HP}$. (b) Selected area diffraction patterns from TEM corresponding to the highest $T_{HP}$, to verify the primary and secondary phases identified with XRD.

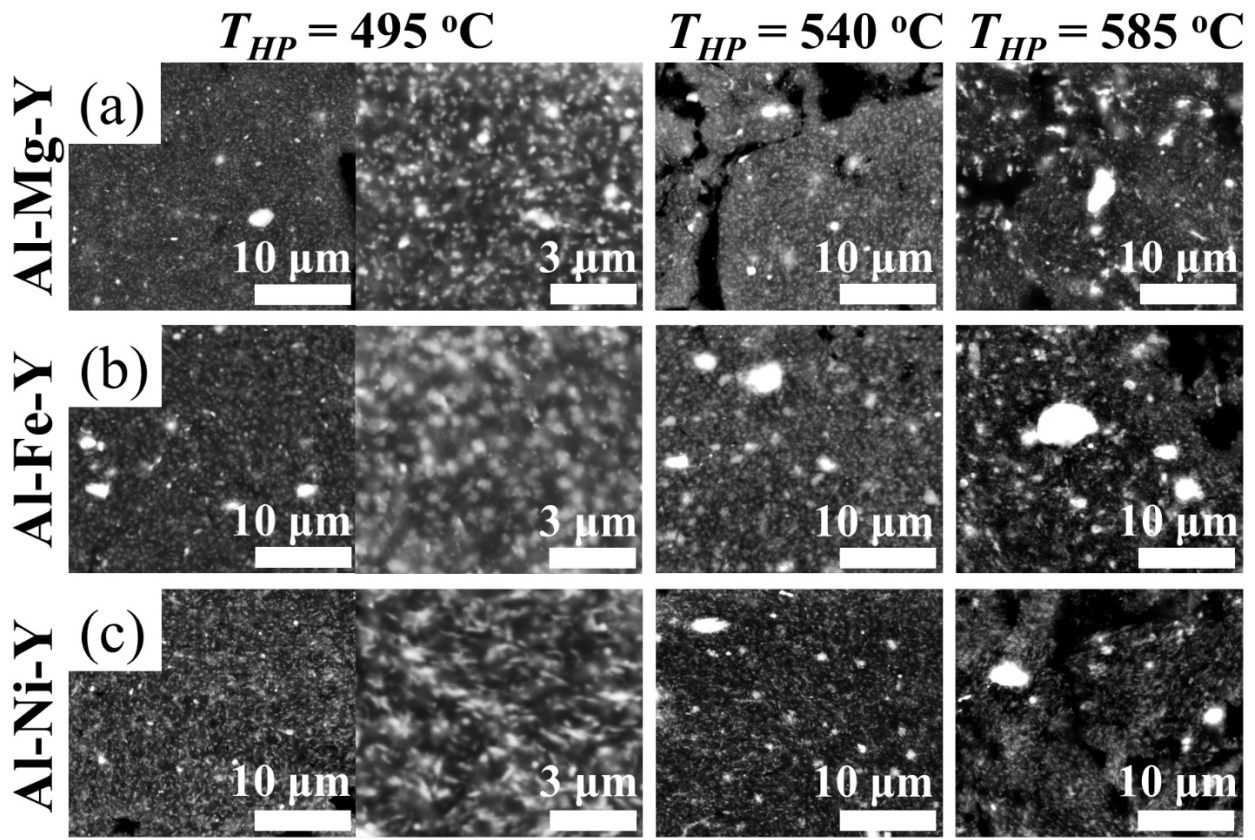

**Figure 12.** (a)-(c) Cross-sectional backscattered electron (BSE) images of intermetallic particles for the three highest $T_{HP}$ values, where only one intermetallic phase dominates and grows in each alloy. For $T_{HP}$ = 495 °C, one magnified BSE image is also presented for each alloy to more clearly show the shape of the intermetallic phases.

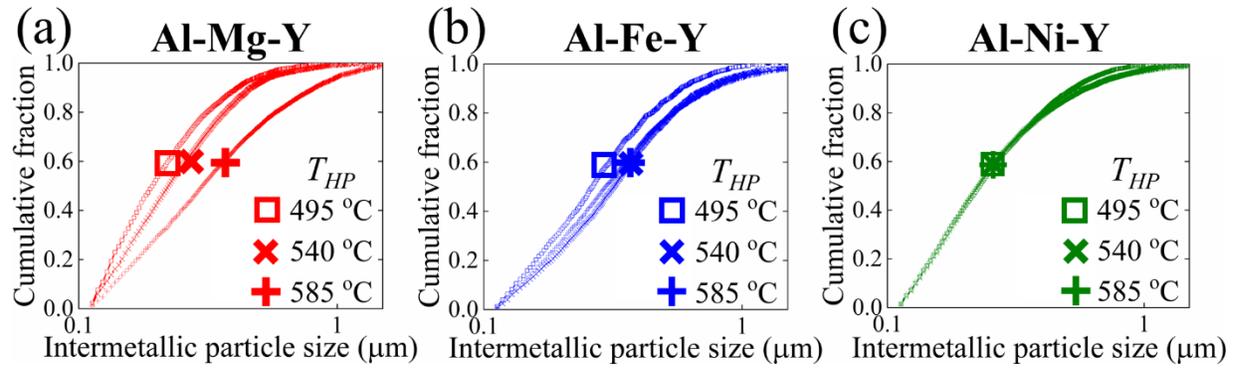

**Figure 13.** Cumulative distribution function of intermetallic particle sizes, as measured by BSE imaging, for (a) Al-Mg-Y, (b) Al-Fe-Y, and (c) Al-Ni-Y. For each alloy, distributions are presented for the three highest $T_{HP}$ values, where only one intermetallic phase dominates and grows in each alloy.

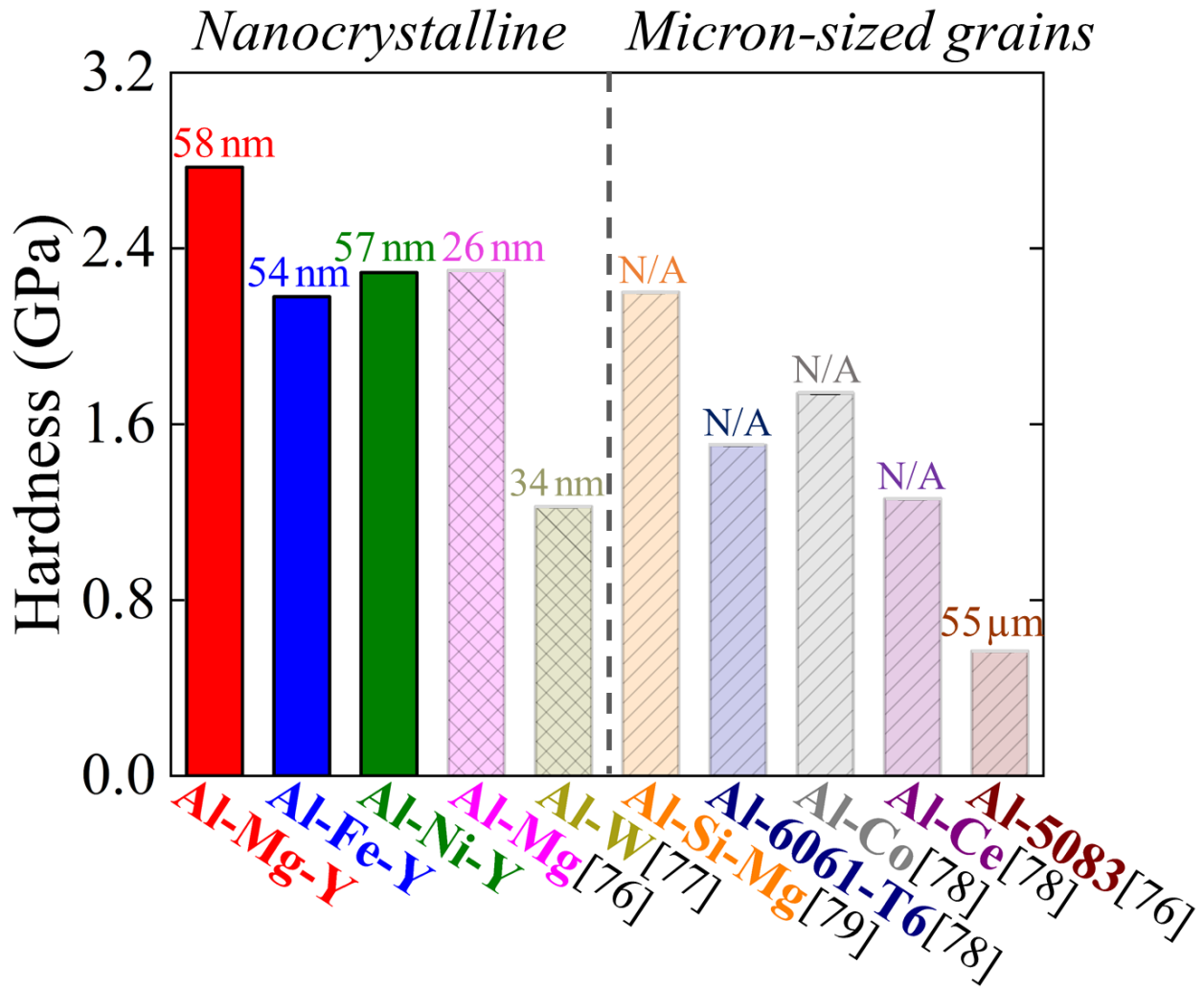

**Figure 14.** Comparison of hardness values of the Al alloys from the present study (solid bars without patterns) with other literature reports (transparent bars with cross or line patterns), where the Al-Mg-Y shows the highest hardness of 2.77 GPa. For the present study, data corresponding to the highest $T_{HP}$ (585 °C) are shown. The value on top of each bar represents the corresponding average grain size, and "N/A" indicates that the grain size was not reported for that alloy in the associated reference.

**Table 1.** Dopant compositions and XRD grain sizes of the Al-rich FCC phase for both as-milled and annealed powders. Two annealing temperatures are investigated, one intermediate (270 °C) and one higher (540 °C). All annealing treatments were followed by water quenching.

| Alloy | EDS dopant concentration (at.%) | | | | XRD grain size (nm) | | |
|---|---|---|---|---|---|---|---|
| | Mg | Fe | Ni | Y | As-milled | 270 °C for 1 h | 540 °C for 1 h |
| Al-Mg-Y | 2.1 | 0 | 0 | 2.2 | 19 | 19 | 39 |
| Al-Fe-Y | 0 | 2.2 | 0 | 2.1 | 11 | 11 | 34 |
| Al-Ni-Y | 0 | 0 | 2.2 | 2.1 | 10 | 10 | 34 |

**Table 2.** Pair-wise mixing enthalpy between the elements investigated in the present study and the atomic radius of each element (shown in parentheses).

| Element | Al <br> ($r_{Al}$ = 1.43 Å [30]) | Mg <br> ($r_{Mg}$ = 1.60 Å [30]) | Fe <br> ($r_{Fe}$ = 1.28 Å [30]) | Ni <br> ($r_{Ni}$ = 1.28 Å [30]) |
|---|---|---|---|---|
| Al | N/A | -2 kJ/mol [39] | -11 kJ/mol [41] | -22 kJ/mol [41] |
| Y <br> ($r_Y$ = 1.80 Å [31]) | -38 kJ/mol [39] | -6 kJ/mol [39] | -1 kJ/mol [40] | -31 kJ/mol [39] |

**Table 3.** Hardness values measured from nanoindentation tests for Al-Mg-Y, Al-Fe-Y, and Al-Ni-Y pellets with the highest hot pressing temperature ($T_{HP}$ = 585 °C) and then naturally cooled down to room temperature.

| Alloy | Hardness (GPa) |
|---|---|
| Al-Mg-Y | 2.77 ± 0.12 |
| Al-Fe-Y | 2.18 ± 0.15 |
| Al-Ni-Y | 2.29 ± 0.16 |